\pdfoutput=1

\documentclass[11pt]{article}

\usepackage[final]{acl}

\usepackage{times}
\usepackage{latexsym}

\usepackage[T1]{fontenc}

\usepackage[utf8]{inputenc}

\usepackage{microtype}

\usepackage{inconsolata}

\usepackage{graphicx}

\usepackage{comment}
\usepackage[whole]{bxcjkjatype}
\usepackage{arydshln}
\usepackage{booktabs}
\usepackage{tcolorbox}
\tcbuselibrary{breakable}
\usepackage{listings}
\usepackage{multirow}

\usepackage{color}

\usepackage{algorithm}
\usepackage{algpseudocode}
\algrenewcommand\algorithmicrequire{\textbf{Input:}}
\algrenewcommand\algorithmicensure{\textbf{Output:}}

\title{Schema-Guided Response Generation using Multi-Frame Dialogue State for Motivational Interviewing Systems}

\author{Jie Zeng \and Yukiko I. Nakano \\
  Seikei University \\
  \texttt{\{jie-zeng, y.nakano\}@st.seikei.ac.jp}
}

\begin{document}
\maketitle
\begin{abstract}

The primary goal of Motivational Interviewing (MI) is to help clients build their own motivation for behavioral change. To support this in dialogue systems, it is essential to guide large language models (LLMs) to generate counselor responses aligned with MI principles. By employing a schema-guided approach, this study proposes a method for updating multi-frame dialogue states and a strategy decision mechanism that dynamically determines the response focus in a manner grounded in MI principles. The proposed method was implemented in a dialogue system on two different datasets and evaluated through a user study. Results showed that the proposed method successfully generates responses aligned with MI principle and frequently asks questions to elicit change talk\footnote{
The code and dataset associated with this study is 
available at \url{https://github.com/IUI-Lab/schema-guided-mi}}.

\end{abstract}

\section{Introduction}
Motivational interviewing (MI)
is defined as ``a collaborative, goal-oriented style of communication 
... designed to strengthen personal motivation for and commitment to a specific goal by eliciting and exploring the person's own reasons for change...'' \cite{miller2012mi_book}.
This definition emphasizes that MI is a goal-directed dialogue aimed at motivating clients to change their behavior and this goal should be achieved by eliciting statements towards change from the client.
Consequently, MI prioritizes understanding the client's perspective rather than talking from the counselor's viewpoint \citep{wu-AnnoMI-23,wu-fs-prompt-ack23}. High-quality MI is therefore characterized by more questioning and reflective listening, whereas unfavorable responses include advising without permission and confronting \citep{perez-rosas-acl2019-highlow}.
\begin{figure}[th]
  \centering
  \includegraphics[keepaspectratio,scale=0.52]{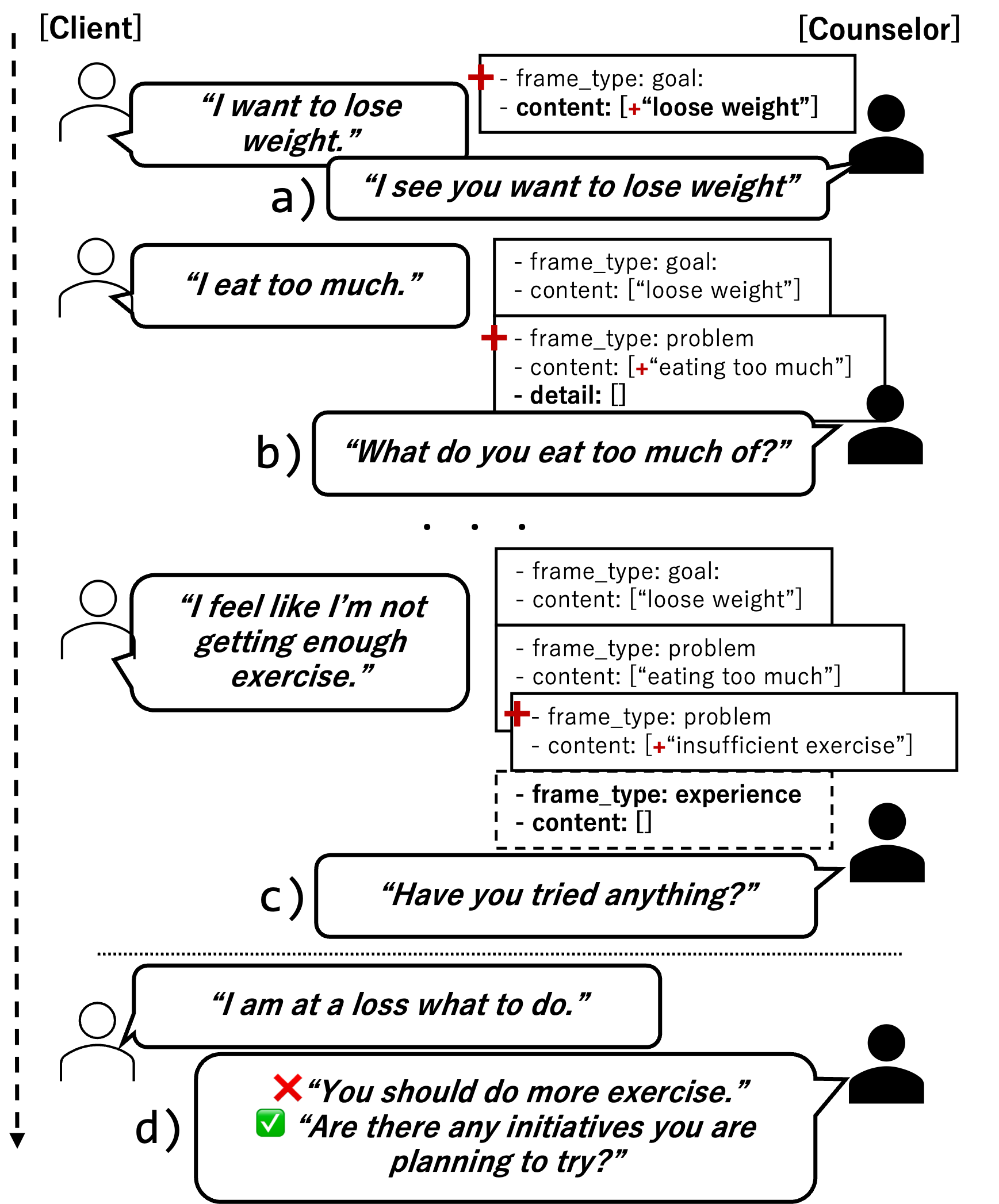}
  \caption {Example of professional MI interaction along with dialogue state. }
  \vspace*{-0.3cm}              
  \label{fig:dialogue_sample}
\end{figure}

Recent advances in large language models (LLMs) enable AI chatbots to address the growing societal need for accessible and personalized support in areas such as mental health, mental well-being, and behavior change (e.g., smoking cessation, alcohol abstinence, and diet) \citep{fitzpatrick_17_cbt, inkster_18_empathy, welch-emnlp20-covid-interview, mousavi-aclws2021-CBT}. Within these applications, MI has emerged as one of the key targets for automated intervention. 

Prior work on AI chatbots for MI has explored prompt-based methods that instruct LLMs to adopt a counselor role \citep{steenstra_iva24,mahmood-tabako-bot-acl25find}. 
Other approaches leveraged LLMs to generate MI-favorable responses, including reflections \citep{shen-MIresponse_sigdial20,shen-KG-MIreflection_ACL22} and advice with permission \citep{welivita-pu-2023-boosting}. More recent studies have proposed methods for exploring various MI topics \citep{yang-ACL25-CAMI} and for learning dialogue strategies from corpora \citep{xie24-MI-strategy}, further promoting MI adherence.
While these previous studies have shown that incorporating richer domain-specific and contextual knowledge can improve MI adherence in counselor response generation, further improvements are still needed.
In particular, overreliance on AI is a critical issue that must be addressed. Prior work on AI ethics shows that LLM-based counseling systems may impose solutions without asking for additional context and lecture users into change \cite{Iftikhar_LLMCounselorViolate_2025}, while HCI research found that people tend to overrely on AI recommendations in decision-making even when those recommendations are incorrect \cite{Buccinca_2021}. These findings highlight the need for techniques that reduce overreliance on AI by prioritizing client autonomy and collaborative dialogue over directive guidance.

To address these issues, this study aims to realize professional MI conversations shown in Figure \ref{fig:dialogue_sample}-a–c. These conversations often involve multiple interrelated topics, 
such as the client's goal (e.g. ``I want to lose weight'') and challenges and obstacles to achieve the goal (e.g. ``I eat too much,'' ``I feel like I'm not getting enough exercise'').
To manage such interactions, we propose a multi-frame dialogue state that maintains multiple streams of dialogue state corresponding to interrelated topics and captures specific and concrete information from the dialogue. 
To further improve toward MI-adherent response generation, we introduce dialogue strategy schemas, grounded in MI principles, that effectively leverage this structured dialogue state.
Specifically, we adopt a schema-guided approach \citep{Rastogi_aaai20_guide_dataset, Li-Stimulus-prompt_NeurIPS23, zeng-sigdial23-question}, in which both dialogue states and dialogue strategies are represented as frames, guiding LLMs toward generating MI-favorable responses. While schema-guided methods have proven effective for dialogue state tracking (DST) with predefined topics, MI requires more flexible dialogue management with dynamically evolving topics. To address this challenge, we propose a method for producing dialogue strategies by leveraging MI knowledge retrieved from an MI corpus.

\noindent{\textbf{Contributions.}}
Employing a schema-guided approach, this study 1) proposes a method for updating dialogue state using multi-frame representation, 2) introduces a mechanism for producing frame-based dialogue strategies that guide LLMs to generate responses aligned with MI principles, and 
3) conducts a user study to evaluate the proposed method by implementing MI dialogue systems on two different datasets, 
demonstrating its effectiveness compared to state-of-the-art approaches 
in terms of MI adherence and active listening.

\section{Related Work}
\subsection{Response Generation in MI}
Research on response generation for MI has addressed two main challenges: generating MI-favorable responses and generating contextually appropriate responses. 

For generating MI-favorable responses, specifically for reflection generation, previous studies proposed fine-tuning and prompt-engineering approaches \citep{wu-fs-prompt-ack23, basar_aclws24_MI-LLM} and also attempted to use knowledge graph (KG) of medical domain \citep{shen-MIresponse_sigdial20, shen-KG-MIreflection_ACL22, deng-acl23-KGEmotion}. There was also a study that generated 
reflection and MI-consistent advice
using rephrasing technique by LLMs \citep{welivita-pu-2023-boosting}. 

Another line of work addresses counselor response generation by considering dialogue state. Early studies relied on rule-based methods, defining client state as precondition and counselor response as the next action \citep{Schulman-MI-rule_AAAI11,sun2025script}. For example, \citet{Park_19_ruleDS_process} developed an MI chatbot that generates questions based on the four-stage MI framework -- engaging, focusing, evoking, and planning \citep{miller2012mi_book}.
More recently, advances in prompt engineering have enabled LLM-based approaches that incorporate dialogue history and MI principles into prompts to guide response generation \citep{steenstra_iva24}. Some methods adopt a two-step prompting process involving dialogue act prediction followed by response generation \citep{sun-coling25-rethinking}, while others generate dialogue strategies--pairs of counselor behaviors and applicable contexts--and encode them in prompt instructions \citep{xie24-MI-strategy}.
Unlike prior works that only consider high-level intent categories and user states, our method improves response generation by extracting more concrete information from the context and organizing it into a structured representation.

\begin{figure*}[ht]
  \centering
  \includegraphics[keepaspectratio,scale=0.50]{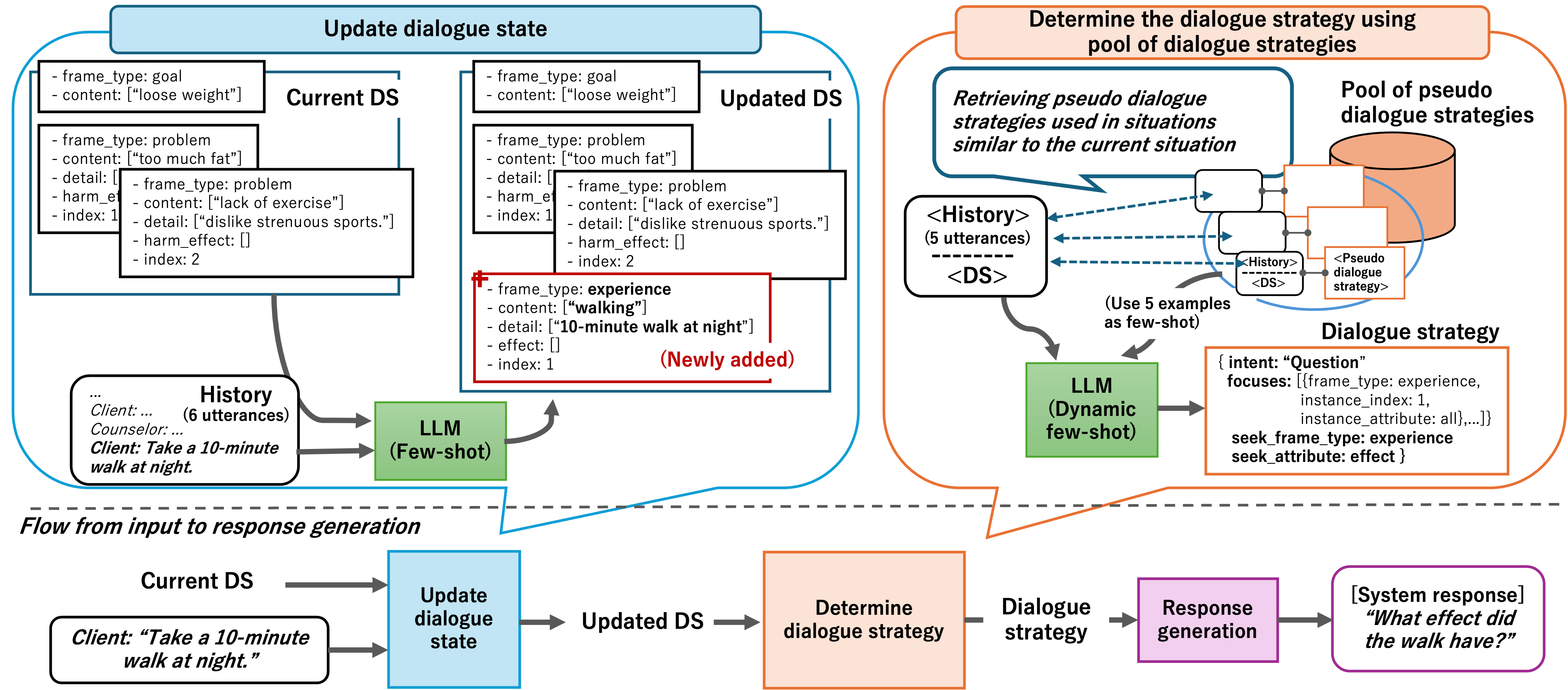}
  \caption{Overall architecture of the proposed mechanism. Content of the dialogue is stored as a dialogue state using Multi-Frame representation (top left: Update dialogue state). The system retrieves dialogue strategies from a pool of pseudo-strategies based on dialogue state similarity and produces a genuine strategy using LLMs (top right: Determine dialogue strategy). Finally, a response is generated based on the determined dialogue strategy. DS denotes Dialogue State.}
  \vspace*{-0.5cm}              
  \label{fig:overview}
\end{figure*}

\subsection{Dialogue Representation and Management}

Previous studies have demonstrated the usefulness of dialogue strategies and policies that contain the intention and semantic content of system responses for effective dialogue management \citep{moon-acl19-opendialkg, Li-Stimulus-prompt_NeurIPS23}.
In particular, in task-oriented dialogue (TOD), the process of updating domain-specific schemas composed of slot-value pairs is known as Dialogue State Tracking (DST) \citep{rastogi_2020schema_DSTC8,Rastogi_aaai20_guide_dataset, chen_emnlp20multi-view,mehri_sigdial2021schema, li-acl24-LLM-DST}.
Recently, methods leveraging prompt engineering with LLMs have been
proposed to perform zero-shot DST and response generation, showing substantial performance \citep{stricker-paroubek-sigdial24-shot, li-acl24-LLM-DST}.

On the other hand, dialogues such as negotiation and emotional support have explicit goals, but the information that needs to be elicited from the user cannot be predefined and often varies depending on the user. 
Generating such dialogues is a challenging task, as dialogue systems require flexible dialogue control. 
To address this, several models have been proposed to learn dialogue policies from annotated dialogue corpora using supervised fine-tuning (SFT) or reinforcement learning (RL) \cite{Deng_Policy_ICLR24, he-dual-planning_acl24}. However, these approaches require large-scale annotated dialogue datasets and model training using a simulator. 
In contrast, this study proposes schema-guided dialogue control that leverages prompt engineering with LLMs, enabling effective dialogue management in low-resource settings using only a small amount of annotated dialogue data.

\section{Methodology}
\label{sec:methodology}
Figure \ref{fig:overview} illustrates the overall architecture of the proposed mechanism and the process of the integrated dialogue system. First, upon receiving user (client) input, the multi-frame dialogue state is updated through a dialogue state tracking mechanism (Section \ref{subsec:update_DS}). Next, using the updated dialogue state and dialogue history as input, a dialogue strategy is generated via a strategy generation module grounded in MI knowledge (Section \ref{subsec:determine_dialogue_strategy}). Finally, based on the generated strategy, a system (counselor) response is produced (Section \ref{subsec:RG}).

\subsection{Multi-Frame Dialogue State}
\label{subsec:update_DS}
\subsubsection{Dialogue State Representation}
\label{subsubsec:design_ds}

With the aim of maintaining MI dialogue states in a structured yet flexible manner, 
each topic is represented as a frame, and the dialogue state is defined as the set of all frames that have emerged during the conversation. This dialogue state provides the foundation for managing the interaction.

Based on an analysis of an MI corpus grounded in MI theory, we identified four schema types---goal, problem, experience, and plan---as templates for frames corresponding to the major topic categories in MI dialogues (see Figure \ref{fig:app_DS_schema}). Every frame must include two core attributes: \textit{frame\_type}, which denotes the type of frame, and \textit{content}, which provides overview phrases describing the frame. In addition, all schema types except for goal include a \textit{detail} attribute that describes more specific information. Some schema types may also contain attributes unique to their schema. For instance, the Problem schema includes \textit{harm\_effect}, which describes the negative consequences of the problem, and \textit{necessity\_to\_improve}, which indicates the perceived need for improvement regarding the problem.
The dialogue segments a--c in Figure \ref{fig:dialogue_sample} show one goal frame and two problem frames.

\subsubsection{Dialogue State Update Mechanism}
\label{subsubsec:update_ds_mechanism}

The left part of Figure \ref{fig:overview} illustrates the flow of updating the dialogue state, which is represented as a set of frames. We leverage an LLM to produce new frames and update the dialogue state.

The LLM takes as input the current dialogue state and a dialogue history consisting of the client's latest utterance and the five preceding ones (six in total). It then generates an updated dialogue state.
We use a few-shot prompting approach (the prompt is shown in Figure \ref{fig:app_extract_DS_prompt}), providing 10 manually constructed input–output examples. These examples were derived from MI dialogue corpus described in Section \ref{subsec:datasets}.

However, the LLM-generated dialogue states sometimes exhibit flaws,
such as omitting frames that were present in the previous dialogue state or redundantly adding new frames, resulting in duplicate values already registered in the dialogue state. 
To address this problem, we implemented a post-processing step that identifies differences between the previous and generated dialogue states and merges the differing parts into the original dialogue state to finalize the update. For details on the prompt and implementation, see Section \ref{app_UpdateDS_detail}.

\subsection{Generating Dialogue Strategies based on MI Knowledge}
\label{subsec:determine_dialogue_strategy}
The dialogue strategy decision mechanism determines the optimal dialogue strategy, which guides the generation of the next counselor response based on the dialogue history and the current dialogue state.

\subsubsection{Dialogue Strategy Representation}
\label{subsubsec:representation_dialogue_strategy}

A dialogue strategy consists of a response intent and a set of attributes with associated values that represent information to be referred in the response. The attributes are defined below.

\noindent{\textbf{intent:} Represents the communicative intent of the response. Five types of intent are defined: four corresponding to core techniques in MI --Question, Affirmation, Reflection, and Summarization -- as well as an additional Other category.}

\noindent{\textbf{focuses:} Specifies the frames in the dialogue state to be focused on in generating the next response. Each focused frame is identified by its frame index, along with optional frame attributes if necessary.}

\noindent{\textbf{seek\_frame\_type, seek\_attribute:} Specifies the type of information to be elicited from the client. The seek\_frame\_type attribute specifies a target frame type (e.g., problem or experience). 
To request the value of a specific attribute in a particular frame, the name of the desired attribute is specified in the seek\_attribute slot.

For example, in Figure \ref{fig:overview}, a counselor strategy to explore the effect of an action taken by the client is represented as:
[intent = ``Question'', focuses = [(references to the relevant dialogue state)], seek\_frame\_type = ``experience'', seek\_attribute = ``effect''].

\begin{algorithm}[t]
\caption{Pseudo Dialogue Strategy Pool}
\label{alg:dialogue_strategy_pool}
{\small

\begin{algorithmic}[1]
\Require dialogue $C$, dialogue state update mechanism $U$, generator LLM $G$, initial turn index $i$

\State StrategyPool $\leftarrow [\,]$
\State $DS \leftarrow \textsc{InitDialogueState}()$

\While{$i < \textsc{TotalTurns}(C)$} 
    \Statex  \textbf{STEP 1) get new data from corpus}
    \If{$\textsc{Speaker}(C,i) \neq \textsc{counselor}$}
        \State $i \leftarrow i + 1$
        \State \textbf{continue}
    \EndIf
    \State $h \leftarrow \textsc{GetTurns}(C, i-5, i-1)$ 
    \Comment {dialogue history}
    \State $target \leftarrow \textsc{GetTurn}(C, i)$
    \Comment{counselor response}

    \Statex \textbf{STEP 2) update dialogue state}
    \State $DS \leftarrow U.\textsc{update}(DS, h, target)$

    \Statex \textbf{STEP 3) generate pseudo dialogue strategy (pds)}
    \State $pds \leftarrow \{\}$
    \State $pds.\textit{intent} \leftarrow \textsc{Intent}(target)$
    \State $pds.\textit{other\_attributes} \leftarrow G.\textsc{GenerateAttrValues}(DS, h, target)$

    \State StrategyPool.\textsc{append}$(pds)$
    \State $i \leftarrow i + 1$
\EndWhile

\State \Return StrategyPool
\end{algorithmic}
}
\end{algorithm}

\subsubsection{Dialogue Strategy Decision Mechanism}
\label{subsubsec:dialogue_strategy_decision_mecanism}

As illustrated in Figure \ref{fig:overview}, the dialogue strategy decision mechanism retrieves dialogue strategies from a pseudo-strategy pool. It selects strategies corresponding to the dialogue state similar to the current one and provides them to the LLM as few-shot examples. Based on these examples, the LLM generates a genuine dialogue strategy for the current context.

\noindent{\textbf{■ Creating a Dialogue Strategy Pool}}

The dialogue strategy pool is a collection of pseudo-strategies that conform to the specification in Section~\ref{subsubsec:representation_dialogue_strategy} and are automatically constructed from an MI dialogue corpus using an LLM. 

Algorithm~\ref{alg:dialogue_strategy_pool} outlines the procedure for generating pseudo dialogue strategies. First, for each counselor response (\textit{target}), the five preceding turns (\textit{dialogue history}) are extracted from the MI dialogue. 
The dialogue state (\textit{DS}) is then updated using both the \textit{target} and the \textit{dialogue history}, according to the dialogue state update mechanism described in Section~\ref{subsubsec:update_ds_mechanism}.
Finally, a pseudo dialogue strategy is constructed by copying the intent of the \textit{target} into the \textit{intent} attribute of the strategy. 
The dialogue history (\textit{h}), the \textit{target}, and the updated dialogue state (\textit{DS}) are then provided to an LLM to generate the remaining attributes, such as  \textit{focuses}, \textit{seek\_frame\_type}, and \textit{seek\_attribute}.

The prompt used for pseudo-strategy generation is shown in Figure~\ref{fig:app_policy_gen_pseudo}. Note that utterance categories are pre-annotated for each counselor utterance in the MI corpus (Section~\ref{subsec:datasets}) and are used as the \textit{intent} of the counselor's target response.
All remaining attributes are generated by the LLM.
Using this method, we generated pseudo dialogue strategy samples from five MI dialogues and stored them in the dialogue strategy pool together with the corresponding dialogue history and dialogue state. (For details on the prompts and implementation, see Section~\ref{app_create_pseudo_strategy}.)

\noindent{\textbf{■ Generating a Dialogue Strategy}}

To determine a genuine dialogue strategy, our method first retrieves several pseudo-strategy samples from a pool, whose associated dialogue states and preceding utterances are most similar to the current dialogue state and the most recent dialogue history.
These retrieved samples are then used as dynamic few-shot examples to prompt an LLM to generate a genuine dialogue strategy.

As illustrated on the right side of Figure~\ref{fig:overview}, the system constructs a text input by concatenating the current dialogue state with the dialogue history comprising the five most recent utterances. This input is encoded using a pre-trained encoder-based language model to obtain a query embedding. Similarly, for each sample in the dialogue strategy pool, we precompute embeddings for each pair of dialogue state and history after concatenating them.
Using the query embedding, we retrieve the top-N samples from the dialogue strategy pool based on cosine similarity (with $N=5$ in our experiments). 

The dialogue states, dialogue histories, and pseudo dialogue strategies of the retrieved samples are then formatted as few-shot examples and provided to the LLM to generate the genuine dialogue strategy. Implementation and prompt details are provided in Figure~\ref{app_policy_gen} and Appendix~\ref{app_determine_strategy}.

\subsection{Response Generation}
\label{subsec:RG}

The response generation process takes a frame-based dialogue strategy (Section~\ref{subsubsec:dialogue_strategy_decision_mecanism}) as input and generates a response using an LLM prompted as shown in Figure~\ref{app_RG_prompt}.

However, LLM-generated dialogue strategies are prone to being logically inconsistent. For example, \textit{Summarization} intent might be combined with an information-seeking slot like seek\_frame\_type = "experience".
To mitigate this issue, we use explicit prompt instructions that prioritize relevant attributes during strategy generation.
Specifically, when the intent is \textit{Reflection} or \textit{Summarization}, the prompt instructs the model to prioritize information represented in the current dialogue state and to generate a response that reflects that content. In contrast, when the intent is \textit{Question} or \textit{Other} and attributes such as \textit{seek\_frame\_type} or \textit{seek\_attribute} are present, the prompt directs the LLM to generate a question that elicits the specified information from the user. See Appendix \ref{app_RG} for prompt and implementation details.

\section{Experiments}

To assess the performance of the proposed method, \textbf{S}chema-\textbf{G}uided \textbf{R}esponse generation for \textbf{MI} (\textbf{SGR-MI}), in real-time user interactions, we implemented the SGR-MI method in a dialogue system, and conducted a user study with 192 participants. We also validated the versatility and domain-independence of the method by testing its effectiveness across two distinct datasets.

\subsection{Baselines}

We compare SGR-MI with three state-of-the-art MI generation methods (1–3) and an ablated version of our model (4) to evaluate the contributions of each component.
1) \textbf{MI Few-shot (MI-FS)} incorporates the dialogue history of the current interaction into the prompt, along with five dialogue samples for few-shot learning, each consisting of 15 utterances by following \citet{wu-fs-prompt-ack23}.
Note that this approach lacks an explicitly structured dialogue state and an explicit mechanism for dialogue strategy determination.
2) \textbf{DIIR} \citep{xie24-MI-strategy} generates responses based on the client's state and dialogue strategies, which are described as natural language inductive rules learned from an MI corpus.
Compared to DIIR, SGR-MI represents the client state with explicit dialogue content and produces strategies that incorporate both frame information from the dialogue state and the specific attributes targeted in the subsequent response.
3) \textbf{CAMI} counselor agent \citep{yang-ACL25-CAMI} possesses client state inference and topic exploration capabilities. However, because CAMI was designed for offline dialogue simulation, it was not suitable for inclusion in our user study for two main reasons. First, its response latency was excessively long, with an average generation per counselor response exceeding 28 seconds (see Appendix \ref{tab:app_chat_basic_stats}), making it impractical to conduct a user study. Second, CAMI requires prior knowledge of the client's goals and profile, which is unavailable in real-time human–agent interactions. Consequently, CAMI was excluded from the user study evaluating live human-agent interaction, and its simulator outputs were used only for the analysis of system responses in Section \ref{subsec:characteristic_sys}.
4) \textbf{SGR-MI w/o Strategy} is an ablated variant of SGR-MI that removes the dialogue strategy generation component. While it uses the same intent categories as SGR-MI, it generates responses solely based on the multi-frame dialogue state mechanism, without referencing MI counselor strategies or identifying the focused information in the current dialogue state.
This variant is used only for the evaluation of system responses in terms of adherence to MI principles in Section \ref{subsubsec:miti_code}.

\subsection{Dtatasets}
\label{subsec:datasets}

We evaluated the versatility and domain-independence of SGR-MI using two datasets. 

\noindent{\textbf{Diet-MI corpus:}} 
Focusing on diet-related concerns, we recorded 48 online MI sessions conducted by professional counselors, each approximately 20 minutes long \cite{nakano22mi_changetalk}.
While the original sessions were conducted in Japanese, we translated the transcripts into English to implement a dialogue system that responds in English for the user study.

\noindent{\textbf{AnnoMI corpus:}} Following previous studies \citep{xie24-MI-strategy,yang-ACL25-CAMI}, we also conducted a user study using AnnoMI dataset \citep{wu-AnnoMI-23}, in which the dialogue content is not restricted to any specific domain, allowing for diverse MI topics.

\subsection{Implementation Details}
\label{subsec:implementation_detail}

To ensure a fair comparison between the proposed and baseline methods, we aligned experimental settings for the following aspects. 
First, we defined five response intents based on core MI skills: Question, Affirm, Reflection, Summarization, and Other. Examples of these intents were included in the few-shot examples for MI-FS. For DIIR, the same five intents were adopted as ground truth response categories for learning dialogue strategies.
Second, to evaluate the effectiveness of acquiring dialogue strategies from a limited number of samples, we randomly selected five sessions from each dataset and used them to construct the dialogue strategy pool in SGR-MI (Section \ref{subsubsec:dialogue_strategy_decision_mecanism}) as well as to learn inductive rules as dialogue strategies in DIIR. 

To ensure that the user study was realistic and effective, we selected models capable of generating responses within 10 seconds on average. As a result, CAMI was excluded from the user study.

We employed GPT-4o (gpt-4o-2024-08-06) as the LLM backbone for all models.
Further details for the datasets and the implementation of each baseline method are provided in Appendix \ref{app:sec_exp_details}.

\subsection{User Study Design}
As described in Section \ref{subsec:datasets}, Diet-MI is a single-domain corpus focused on diet, whereas AnnoMI is a multi-domain corpus covering diverse MI topics. Because the dialogue systems were implemented based on each corpus, we treat corpus choice as an experimental condition, resulting in a single-domain (Diet-MI) and a multi-domain (AnnoMI) condition.
Consequently, the user study employed a $2 \times 4$ between-subjects factorial design, with domain (single, multi) and model (MI-FS, DIIR, SGR-MI, SGR-MI w/o Strategy) as the primary factors.
Participants were recruited via crowdsourcing, with 24 participants assigned to each of the eight conditions, resulting in a total of 192 participants.

Participants engaged in a text-based interaction with the dialogue system via a web interface. In the single-domain (Diet-MI) sessions, participants were instructed to discuss diet-related concerns. In the multi-domain (AnnoMI) sessions, participants were free to discuss any aspect of their daily lives that they wished to change. 
Participants were required to produce at least 20 inputs (one input per utterance), resulting in dialogues with at least 40 utterances.
This requirement was introduced because real-world MI sessions typically last 30–60 minutes \citep{rubak2005motivational}, suggesting that a certain interaction length is necessary for a meaningful MI process.
After reaching this minimum, participants were free to end the conversation whenever they chose.
Following the dialogue session, participants completed a 16-question survey. The details of the experiments are presented in Appendix \ref{app:subsec_details_user_study}.

\section{Evaluation Results}
\subsection{Analysis of System Response}
\label{subsec:characteristic_sys}

This section analyzes system responses from two perspectives: adherence to MI principles and question types to elicit user thoughts toward change.

\subsubsection{Analysis of Adherence to MI principles }
\label{subsubsec:miti_code}

System responses were evaluated using the Motivational Interviewing Treatment Integrity (MITI) framework, which assesses how well counselor behavior aligns with MI principles and skills. 
Counselor utterances were classified at the sentence level into 11 categories: the ten behavior codes defined in MITI 4 \citep{moyers_16_MITI4}, and an additional ``not applicable (n/a)'' category.
To ensure accurate sentence classification, we relied on manual annotation by human annotators rather than automatic labeling.
Four trained annotators conducted the MITI annotation, achieving a Fleiss's $\kappa$ of 0.60, indicating moderate inter-rater agreement. 
Based on the MITI annotations, we computed six metrics to evaluate counselor behaviors in MI.

\noindent{\textbf{Complex Reflection Ratio (\%CR)}} The proportion of Complex Reflections among Reflections. 
\noindent{\textbf{Reflection-to-Question Ratio (R/Q)}} The ratio of Reflections to Questions by the counselor. \\
\noindent{\textbf{MI Non-Adherent Ratio (\%MINA)}} 
The ratio of MI non-adherent responses (Persuasion and Confrontation) to all counselor sentences. \\
\noindent{\textbf{Percent of Active Listening Behaviors (\%AL)}} The ratio of the sum of Questions and Reflections to all counselor sentences. \\
\noindent{\textbf{Percent of Non-authoritative Behaviors (\%NA)}} 
The ratio of sentences that are not confront, advice (persuade with or without permission), or give information, among all counselor sentences \citep{xie24-MI-strategy}. This metric quantifies the counselor's tendency to avoid information-providing behaviors. \\
\noindent{\textbf{Percent of Therapist Talk Time (\%TTT)}}
The ratio of counselor talk time to total talk time (counselor and client combined) \citep{yang-ACL25-CAMI}. We approximate talk time using word counts.

\begin{table}[]
\resizebox{1\linewidth}{!}{%

\begin{tabular}{lrrrrrr}
\toprule

                                   & \%CR$\uparrow$      & R/Q$\uparrow$        & \%MINA$\downarrow$ & \%AL$\uparrow$     & \%NA$\uparrow$      & \%TTT$\downarrow$ \\
\midrule

\multicolumn{7}{l}{\colorbox{gray!30}{Single-domain (Diet-MI)} } \\
MI-FS                              & \textbf{43.2}        & 0.94                & 9.7                & 14.6                & 33.3                & 73.6 \\
DIIR                               & 39.7                 & \textbf{2.01}$\dag$ & 5.9                & 54.1                & 75.7                & 68.8 \\
SGR-MI                             & 34.8$\dag$           & 0.98                & \textbf{0.0}$\dag$ & \textbf{85.6}$\dag$ & \textbf{95.6}$\dag$ & \textbf{59.6}$\dag$  \\ 
\shortstack{SGR-MI w/o\\Strategy}  & 38.5                 & 0.27                & 17.1               & 41.2                & 55.7                & 60.5 \\
\midrule
GT          & 34.9                 & 2.38                 & 0.1                & 69.9                & 95.1                &  55.2\\
\addlinespace[3mm]

\multicolumn{7}{l}{\colorbox{gray!30}{Multi-domain (AnnoMI)} } \\
MI-FS                              & \textbf{59.3}$\dag$  & 1.28                & 9.9                & 9.1                 & 36.8                 & 69.8 \\
DIIR                               & 46.7                 & \textbf{1.43}$\dag$ & 5.8                & 31.4                & 63.6                 & 69.5 \\
SGR-MI                             & 36.3                 & 0.91                & \textbf{0.0}$\dag$ & \textbf{87.5}       & \textbf{94.7}        & 59.6 \\
\shortstack{SGR-MI w/o\\Strategy}  & 58.7                 & 0.26                & 10.3               & 51.2$\dag$          & 70.2$\dag$           & \textbf{56.1}$\dag$ \\
\midrule
GT                                 & 69.7                 & 1.51                & 1.4                & 59.0                & 81.7                 & 52.7\\
\addlinespace[3mm]

\multicolumn{7}{l}{\colorbox{gray!30}{Combined (Diet-MI+AnnoMI)} } \\
MI-FS                              & \textbf{50.4}$\dag$  & 1.06                & 9.8                & 11.7                 & 35.1                 & 71.5 \\
DIIR                               & 42.2                 & \textbf{1.76}$\dag$ & 5.8                & 42.4                 & 69.4                 & 69.2 \\
SGR-MI                             & 35.5                 & 0.94                & \textbf{0.0}$\dag$ & \textbf{86.6}        & \textbf{95.1}$\dag$  & 59.6 \\
\shortstack{SGR-MI w/o\\Strategy}  & 49.3                 & 0.27                & 13.7               & 46.1$\dag$           & 62.8                 & 58.3$\dag$ \\
\midrule
GT                                 & 52.3                 & 1.94                & 0.7                & 64.5                 & 88.0                 & 54.0 \\
\midrule
\addlinespace[2mm]
\shortstack{CAMI-\\simulator}      & 40.0                 & 0.46                & 18.3               & 24.9                 & 47.8                 & \textbf{49.2} \\

\bottomrule
\end{tabular}

}

\caption{Evaluation of response appropriateness in MI principles. \textbf{Bold} and $\dag$ indicate best performance and values closest to ground truth (GT), respectively. GT values for Diet-MI are computed from annotations of seven Diet-MI dialogues, while values for AnnoMI are taken from \citet{xie24-MI-strategy}. CAMI scores are based on 12 simulated sessions, not from user studies.}
\label{tab:miti_mesure}
\vspace*{-0.5cm}
\end{table}

Table \ref{tab:miti_mesure} shows the results of the single-domain (Diet-MI) and multi-domain (AnnoMI) experiments, as well as their combined results, together with the ground truth (GT) computed from analyses of the MI corpora. 
Note that the "CAMI-simulator" entry reports an analysis of 12 simulated counselor–client dialogue sessions, following \citet{yang-ACL25-CAMI}, rather than results from a human user study.
As the results are consistent across the single-domain and multi-domain experiments, we use the combined results in the following discussion.

Regarding \textbf{\%CR} and \textbf{\%R/Q}, SGR-MI underperformed the baselines, suggesting a higher frequency of questions than reflections in SGR-MI.
For \textbf{\%MI Non-Adherent (\%MINA)}, 
SGR-MI achieved a perfect score of 0.0\%, indicating that it consistently avoids generating responses that violate MI principles, whereas the baseline methods produced MI-inconsistent responses.
Notably, the high \%MINA ratio in SGR-MI w/o Strategy highlights that the strategy determination component is essential alongside the dialogue state mechanism for generating responses that adhere to MI principles.
SGR-MI also achieved the highest \textbf{\%AL} score, suggesting a superior ability to employ active listening behaviors that elicit users' own thoughts. Furthermore, the \textbf{\%NA} metric reveals that SGR-MI rarely produced information-providing behaviors, maintaining higher adherence to MI principles compared to the other methods.
Regarding the \textbf{\%TTT} metric, although the CAMI-simulator achieved the best score, this result is influenced by the behavior of the client simulator, which produced approximately three times more words per utterance than human clients in our user study. Therefore, the balanced talk time observed in simulation may not be generalizable to interactions with human users.
Among the user study conditions, SGR-MI performed comparably to SGR-MI w/o Strategy, which achieved the best \textbf{\%TTT} score.
\begin{table}[]
\resizebox{1\linewidth}{!}{%
\begin{tabular}{lrrr}
\toprule
 & Q Rate & \multicolumn{2}{c}{\shortstack{Change Talk Eliciting\\Question Ratio}} \\
\cmidrule(lr){3-4}
 &  & ev-* & pl-* \\ 
\midrule
MI-FS & 0.057 & 0.005 & 0.006 \\
DIIR  & 0.153 & 0.017 & 0.021 \\
SGR-MI  & \textbf{0.446} & \textbf{0.079} & \textbf{0.097} \\
\midrule
CAMI-simulator & 0.170 & 0.003 & 0.003 \\
\bottomrule
\end{tabular}
}

\caption{Proportions of question categories. \mbox{ev-*} denotes the ratio of evoking questions to all system sentences, \mbox{pl-*} denotes the ratio of planning questions. CAMI-simulator scores were derived from 12 dialogue sessions using the simulator by \citet{yang-ACL25-CAMI}. Tukey's HSD test revealed that SGR-MI significantly outperformed MI-FS and DIIR on all metrics ($p<.01$).}
\label{tab:q_detail}
\vspace*{-0.5cm}              
\end{table}

\begin{figure*}[h]
  \centering
  \includegraphics[keepaspectratio,scale=0.50]{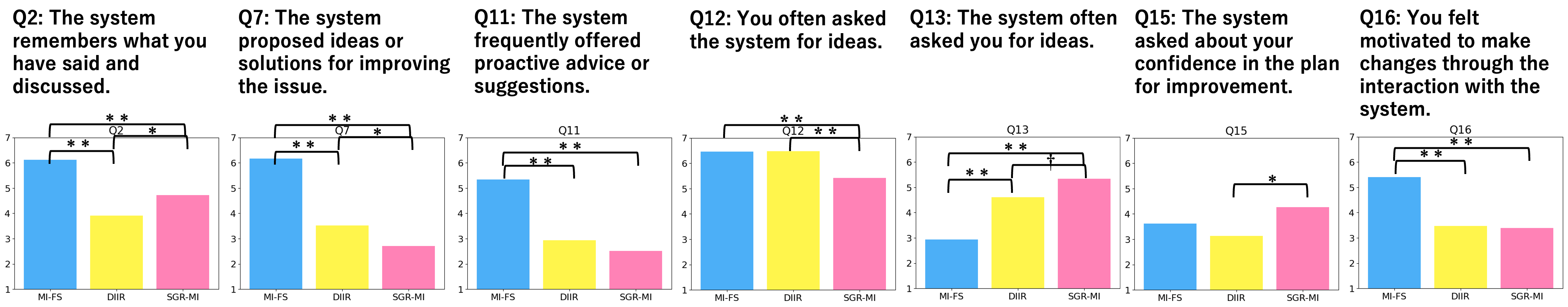}         
  \caption{Main subjective evaluation results. The questionnaire was rated on a 7-point Likert scale (1 = strongly disagree, 7 = strongly agree); p-values were computed using Tukey's HSD.
   ($\dagger:p<.1,\:{*}:p<.05,\:{*}{*}:p<.01$)}
  \vspace*{-0.5cm}              
  \label{fig:enquete_main_ret}
\end{figure*}

\subsubsection{Analysis of Question Types}

The primary objective of MI is to elicit Change Talk (CT), which refers to client utterances directed toward their goals.
In this section, we evaluate the effectiveness of SGR-MI in generating questions designed to facilitate CT. Specifically, we analyze and categorize system-generated questions into evoking and planning stages \citep{miller2012mi_book}.
Following the framework by \citet{miller2012mi_book}, we defined evoking questions as those that ask about the user's desire, ability, reasons, and need for change, with the goal of eliciting intentions to change. Planning questions were defined as those that elicit willingness, commitment, and concrete steps or preparation for change, thereby facilitating the creation of actionable plans for change. We defined nine subcategories for these questions (Appendix \ref{app:annotation_detail}). The same four annotators who conducted the MITI coding annotated these questions, achieving moderate inter-rater reliability (Fleiss' $\kappa = 0.60$). Full frequency results for all categories are reported in Table \ref{tab:app_q_category_full}.

Table \ref{tab:q_detail} reports the Q Rate (the proportion of questions among all system sentences) and the Change Talk Eliciting Question Ratio (the proportion of evoking and planning questions among all system sentences). 
We conducted Tukey's HSD tests to compare MI-FS, DIIR, and SGR-MI. The results show that SGR-MI achieved a significantly higher proportion of change talk eliciting questions than both MI-FS and DIIR, and also outperformed the CAMI-simulator. These findings indicate that the proposed method is more actively seeks to elicit CT than the baselines.

\subsection{Analysis of Client Behavior and Impression}
This section analyzes client behavior and perception under three conditions (MI-FS, DIIR, and SGR-MI) from two perspectives: the elicitation of change talk and subjective impressions of the interaction.

\begin{table}[]
\centering
\resizebox{!}{70pt}{

\begin{tabular}{lrr}
\toprule
   & \shortstack{CT Utterances\\per Session} & Triggered by Advice  \\
\midrule
\multicolumn{3}{l}{\colorbox{gray!30}{Single-domain (Diet-MI)} } \\
MI-FS & 1.96 & 78.7\% \\
DIIR & 2.42 & 25.9\% \\
SGR-MI & 3.96 & 3.2\%\\
GT  & 21.7 & 1.3\%\\
\addlinespace[2mm]
\multicolumn{3}{l}{\colorbox{gray!30}{Multi-domain (AnnoMI)} } \\
MI-FS & 3.88  & 82.8\%\\
DIIR  & 1.96 & 44.7\%\\
SGR-MI & 3.88 & 1.1\% \\
GT  & 31.5 & 1.2\%\\

\bottomrule
\end{tabular}

}

\caption{Client Change Talk (CT) frequency and its advice-triggered ratio.}
\label{tab:CT_trigger}
\vspace*{-0.5cm}
\end{table}

\subsubsection{Analysis of Client Change Talk}
We analyzed client CT in dialogues obtained from our user study as well as professional counselor dialogues (GT) using the Manual for the Motivational Interviewing Skill Code \citep{miller2003misc}. 
As both the Diet-MI and AnnoMI datasets include annotations for client utterance categories, we used these annotations to identify the ground truth (GT) for CT.
Because conversations in Diet-MI are approximately 20 minutes long, we selected from the AnnoMI dataset only sessions with at least 150 utterances (averaging 23 minutes) for analysis.

Table \ref{tab:CT_trigger} shows the average number of CT utterances per session for each condition, along with the percentage of CT triggered by counselor "advice".
Regarding the number of CT utterances, GT sessions contain more than 20 occurrences per session. It should be noted that this higher frequency in GT is likely because it is based on transcriptions of spoken dialogue, whereas the other conditions involve text-based interactions. Among the system conditions, SGR-MI elicited CT with a high frequency, averaging more than three occurrences per session.

Furthermore, the ratio of CT "Triggered by Advice" for SGR-MI was substantially lower than those for DIIR and MI-FS, showing values closer to GT. These results suggest that in professional MI, client CT tends to originate from the client's own reflection and internal thought processes rather than from advice provided by the counselor. SGR-MI successfully replicates this tendency, effectively evoking spontaneous CT. In contrast, MI-FS and DIIR may hinder the elicitation of change talk by leading with counselor-side advice, suggesting that these dialogues do not sufficiently adhere to the core principles of MI.

\subsubsection{Subjective Impression}

We analyzed a 16-item post-interaction questionnaire from the user study, comparing users' subjective impressions across the three systems (SGR-MI, MI-FS and DIIR).
Figure \ref{fig:enquete_main_ret} shows the results for which significant differences between methods were found using Tukey's HSD tests. 
As there were no substantial differences between single-domain and multi-domain experiments, we report the aggregated results.
The complete list of questionnaire items and rating scores are in Appendix \ref{app:subsec_human_evaluation}.

Among the three methods, MI-FS was perceived as providing practical solutions (Q7) and direct advice (Q11), leading users to perceive it as the system that most obediently followed (Q2) and responded to their requests. 
Consequently, users reported the highest motivation when interacting with MI-FS (Q16). This exhibits a typical case of AI overreliance \cite{Buccinca_2021}. Interaction in the MI-FS condition is characterized by a repetitive loop of user advice-seeking and system advice-giving. This aligns with concerns that LLM-based counseling systems may overproduce directive guidance and risk inappropriate therapeutic intervention \cite{Iftikhar_LLMCounselorViolate_2025}. 
\citet{Buccinca_2021} showed that \textit{cognitive forcing} significantly mitigates overreliance on AI. In our context, cognitive forcing corresponds to requiring users (clients) to engage in effortful self-reflection and to exercise autonomy in deliberate thought toward behavioral change. They also reported that such systems receive the least favorable subjective ratings. 
Our results replicate this trade-off exactly: users rated DIIR and SGR-MI less favorably than MI-FS.
Unlike MI-FS, both DIIR and SGR-MI were perceived to avoid proactive suggestions. However, DIIR was perceived as providing ideas or solutions more frequently than SGR-MI (Q7), which in turn prompted users to seek suggestions from DIIR more often (Q12).

SGR-MI was perceived as most frequently soliciting users' ideas (Q13), suggesting that it encourages self-reflection and the externalization of thoughts by frequently prompting users to articulate their own ideas. This is further supported by the highest active listening scores (\%AL) observed in Section~\ref{subsubsec:miti_code}. Furthermore, compared to DIIR, SGR-MI proactively assesses users' confidence in their plans for change (Q15). This effect is enabled by SGR-MI's underlying architecture, which leverages schemas to structure information to be elicited as attributes and maintains them within a frame-based dialogue state.

\section{Conclusion}

In this study, we proposed a dialogue system aiming to guide the user's deliberation and behavioral change through Motivational Interviewing (MI). The system treats various topics discussed in the dialogue as distinct frames and manages them collectively as the dialogue state. The dialogue system also incorporates a mechanism for selecting appropriate dialogue strategies by referencing professional counselors' strategies according to the dialogue context. 
Through a user study comparing our method with existing approaches, 
we demonstrated that our system reduces the frequency of MI-non-adherent behaviors while increasing active listening behaviors. Moreover, by asking questions to elicit user thoughts, the system facilitates Change Talk originating from the users' own ideas, rather than external advice.

\section*{Limitations}
In this study, we defined and represented a finite set of frame types and attributes based on the principles of MI and observations of actual dialogue data. We also utilized four categories to represent the intent of system responses. However, the validity of the representation of the dialogue state handled in this study is not sufficiently verified for the frames that constitute the dialogue state, and the types of attributes could be extended through automatic acquisition of attributes \citep{finch24-SSI}. In addition, counselor responses often convey a wider range of intentions, suggesting the need for further expansion of dialogue strategy.

Moreover, the experiments conducted in this study involved interactions with participants that were completed in a single day, and the results were based on only one session of dialogue. Additionally, since participants were recruited through crowdsourcing, they may have prioritized task completion over genuinely engaging with the consultation content. Given this short-term experimental setup, the effectiveness of our approach from a long-term perspective remains unclear. To achieve the fundamental goal of MI, which is to promote behavioral change, it is essential to gradually enhance the client's intrinsic motivation through sustained and repeated interactions.
Consequently, future investigations necessitate empirical validation through long-term dialogues and the design of dialogue systems that account for extended interactions.

\section*{Ethical Considerations}
This section discusses the ethics relevant to this study.

\subsection*{Ethical Review}
Before conducting the experiments, we consulted with the ethics board of our institution regarding the collection of MI dialogue data from general participants and obtained their approval.

\subsection*{Risks and User Perception}
We developed a dialogue system that plays the role of a counselor in MI dialogues. 
In our user study, participants reported feeling more motivated when interacting with the baseline few-shot LLM system (MI-FS) than with either our proposed method or the DIIR baseline. While this result raises questions regarding the immediate perception of effectiveness given MI's goal of ``motivating users,'' it highlights a critical ethical issue regarding user interaction with generative AI.

\citet{wu-fs-prompt-ack23} reported that laypeople have significantly more difficulty distinguishing between human and AI-generated responses when the model is highly fluent (e.g., GPT-3), whereas experts focus on content appropriateness and are less susceptible to fluency bias. This suggests that laypeople's evaluations may be heavily influenced by response fluency rather than therapeutic appropriateness. In this context, relying solely on fluency-driven interactions poses a risk of over-trust or overreliance on AI—a growing concern in AI ethics. 
This is particularly critical given the risks identified by \cite{Iftikhar_LLMCounselorViolate_2025} regarding the tendency for LLM-based counseling to provide excessive advice and the potential for inappropriate therapeutic interventions.
Our approach seeks to counter this by prioritizing MI consistency and encouraging users to think for themselves, rather than passively accepting fluent but potentially superficial advice. We believe that this contributes to a more ethically responsible use of LLMs in mental health support, even if it results in lower subjective satisfaction scores from lay users.

Moreover, there is a possibility that the generation process using the language model we employed includes inappropriate advice, unsolicited guidance, hallucinations, or responses that are not aligned with MI principles. Furthermore, the system has not been validated by MI experts. Therefore, our system, and AI more broadly, is not at a stage where it can replace human MI counselors.

Accordingly, developers of MI systems must recognize the potential harms and risks involved. It is also essential to inform users who interact with such systems of these risks in advance and obtain their informed consent.

\subsection*{Information and Consent}
The user study was conducted via crowdsourcing. Workers who applied for the task were shown a disclaimer on the instruction page before interacting with the MI systems, stating that the experiment would involve a dialogue with an AI system and that the system's responses might contain errors or inappropriate content. Their consent was obtained at this stage. Additionally, on the questionnaire page following each system session, participants were asked to consent to the use of the collected data solely for research purposes. Furthermore, participants were provided with an opt-out option if they did not wish their dialogue data to be shared externally, even within the bounds of research use.

\section*{Acknowledgments}
This work was supported by JST CREST, Japan, Grant Number JPMJCR2563 and JSPS KAKENHI Grant Number JP24K02984.

\bibliography{custom}

\appendix

\section{Details of the Proposed Methods}
We present below the prompts and generation details used in our proposed system.

\subsection{Update Dialogue State}
\label{app_UpdateDS_detail}
Figure \ref{fig:app_DS_schema} illustrates the structure of the dialogue states targeted in this study, represented in Pydantic format. The dialogue state comprises four frame types: \textit{GoalAndIdealFrame}, \textit{ProblemAndTroubleFrame}, \textit{ExperienceFrame}, and \textit{ImprovementPlanFrame}. These are integrated into a single class named \textit{Dialogue\_State}. By utilizing this structural definition in conjunction with the function calling capability of the LLM, we can obtain outputs that adhere to the intended dialogue state structure. The prompt used for extracting dialogue states is shown in Figure \ref{fig:app_extract_DS_prompt}.
To improve information extraction accuracy, we annotate each MI session conducted by a professional with the following elements: the dialogue state immediately before the interaction (<Current\_DS>), the most recent utterances from each of the counselor and the client (Utterance), and the updated dialogue state after the interaction (<Updated\_DS>). A total of 10 such annotated samples are prepared and presented in the Examples section.
Information necessary for the current dialogue state extraction is given as the “User Message”, while the <Current\_DS> field contains the immediately preceding dialogue state. The field embeds the recent dialogue history, consisting of the six most recent utterances between the counselor and the client, formatted as ``Counselor:'' and ``Client:''. A sampling temperature of 0 was used during generation to maximize determinism in the output.

\subsection{Creating Pseudo Dialogue Strategy Samples}
\label{app_create_pseudo_strategy}
This section details the generation process of pseudo dialogue strategy samples used as few-shot examples during dialogue strategy generation. Figure \ref{fig:app_policy_gen_pseudo} presents the prompt used to generate pseudo dialogue strategies and the structure of the resulting strategies in Pydantic format. To identify the dialogue strategy employed in a counselor's surface-level response without relying on manual annotations, we utilize utterances from professional MI dialogue datasets. Specifically, we input the counselor's response (<Counselor\_Response>), the preceding five utterances as the dialogue history (<History>), and the dialogue state derived from this history (<Dialogue\_State>) into the LLM, which then generates a pseudo dialogue strategy corresponding to the counselor's response.

Since the system at inference time selects a dialogue strategy that includes an \textit{intent} attribute, it is desirable for the pseudo dialogue strategy samples to also contain this attribute. Fortunately, the counselor responses in the MI dataset are already annotated with response categories, so we directly use these annotations as the \textit{intent} value. Therefore, the generated pseudo dialogue strategies exclude the \textit{intent} attribute from their structure. The sampling temperature during generation was set to 0.

Using the method described above, in the single-domain (Diet-MI), we generated 467 pseudo dialogue strategy samples from five dialogue sessions in the Diet-MI corpus, while in the multi-domain (AnnoMI), 919 samples were derived from five sessions in the AnnoMI corpus.

\subsection{Determine Dialogue Strategy}
\label{app_determine_strategy}
Figure \ref{app_policy_gen} presents the prompt used to determine the dialogue strategy, as well as the schema (in Pydantic format) of the output dialogue strategy. Unlike the pseudo dialogue strategy generation shown in Figure \ref{fig:app_policy_gen_pseudo}, the structure generated by the LLM includes an additional intent attribute. Accordingly, the DialogueStrategy in Figure \ref{app_policy_gen} contains this intent field.

In the prompt, five similar examples are retrieved from a pool of pseudo dialogue strategies and presented in the ``Examples'' section as references for strategies used in similar dialogue situations. Following this, the current dialogue context is embedded into the prompt: five utterances between the client and counselor are included in the <History> section, and the corresponding dialogue state is provided in <Dialogue\_State>.

To retrieve dialogue situations similar to the current one from the pool of pseudo-dialogue strategies, we used the following template: \textit{``\#\#\# History\textbackslash n<History>\textbackslash n\textbackslash n\#\#\# Dialogue\_State\textbackslash n<Dialogue\_State>''}.
The dialogue history of five utterances is embedded in <History> and the dialogue state is embedded in <Dialogue\_State>. 
To obtain text embeddings, we employ a RoBERTa-based XLM-R Longformer model\footnote{\url{https://huggingface.co/markussagen/xlm-roberta-longformer-base-4096}}, which is capable of handling long input sequences comprising both dialogue state and history. XLM-R Longformer is released under the Apache 2.0 license and was adopted in this study to obtain multilingual embedding representations in accordance with its intended purpose.
The sampling temperature was set to 0 when generating the dialogue strategy.

\subsection{Response Generation}
\label{app_RG}
Figure \ref{app_RG_prompt} shows the prompt used for system response generation.
In the <History> section, we embed the most recent six utterances from the dialogue history between the client and the counselor, including the client's latest utterance. The <Dialogue\_State> section contains the dialogue state, along with the pre-determined dialogue strategy elements: \textit{intent, focuses, seek\_frame\_type, and seek\_attribute}, which are inserted into their respective placeholders. The sampling temperature during generation was set to 1.0.

\section{Details of Experiments}
\label{app:sec_exp_details}

\subsection{Dataset}
For both the proposed method and the baselines, we utilized subsets of dialogue datasets involving professional MI counselors. In the single-domain sessions, we employed the dataset collected in our previous work \citep{nakano22mi_changetalk}, while in the multi-domain sessions, we utilized the AnnoMI dataset \citep{wu-AnnoMI-23}. This section describes the dataset from our previous work.

We collected approximately 20-minute counseling dialogues in Japanese, conducted by four professional MI counselors with lay individuals acting as clients. The conversations focused on concerns related to diet and eating habits. In total, we collected approximately 50 sessions; these were transcribed, and each utterance was annotated based on the Motivational Interviewing Skill Code (MISC) \citep{miller2003misc}. Upon publication of this paper, we plan to release the transcripts for the dialogues where consent for data publication was obtained from both counselors and clients.
Although the original dataset is in Japanese, the experiments in this study were conducted in English. Therefore, we used English translations of the Japanese transcripts generated via the Google Cloud Translation API.

\subsection{Common Experimental Settings}
\label{app:subsec_common_exp_settings}
We adopted GPT-4o (gpt-4o-2024-08-06) as the LLM for all experimental conditions. 
In our proposed method, we leveraged its function calling capability to handle structured data, such as dialogue states, and to generate outputs in a specific structure.

To control the response length, we added instructions to the response generation prompts of all systems to limit the output to approximately two sentences. However, strict control was not always achieved, and the actual output sentence lengths varied. Please refer to Table \ref{tab:app_chat_basic_stats} for the specific generation times.

\subsection{Baseline Details}
\label{subsec:app_baseline_details}
Details of the two baselines used in the experiment are provided below.

\noindent{\textbf{MI Few-shot (MI-FS):}}\\
The prompt used for MI Few-shot (MI-FS) is shown in Figure~\ref{fig:app_MI_fs_prompt}. Following the setup of \citet{wu-fs-prompt-ack23}, we construct five samples of professional MI dialogues, each consisting of a dialogue history of fifteen utterances between a client and a counselor, followed by the counselor's response. These samples are inserted into the prompt without randomizing their order as few-shot examples. In the same format as the samples, the current dialogue history is appended, and the system response is generated using the LLM.
When selecting the dialogue samples for few-shot prompting, we ensured that the counselor's responses included the same type of intent (\textit{Question, Affirmation, Reflection, or Summarization}) as required by the proposed system.
The temperature for generation was set to 1.0, following \citet{wu-fs-prompt-ack23}.

\noindent{\textbf{Dialogue strategy Learning via Inductive Reasoning (DIIR):}}\\
DIIR, proposed by \citet{xie24-MI-strategy}, necessitates the prior construction of strategy rules mapping client dialogue states to counselor responses. Therefore, for each experiment, we constructed these strategy rules using the same subset of dialogue data employed to generate the strategy pool for our proposed method. In the single-domain (Diet-MI), we generated 307 strategy rules from five dialogue sessions in Diet-MI corpus, while in the multi-domain (AnnoMI), 400 strategy rules were derived from five sessions in the AnnoMI corpus.

For client dialogue state inference in DIIR, we used the most recent six utterances, including the latest client input, as the dialogue history input for the LLM; this setting aligns with that of the proposed method. We adopted the default hyperparameters specified in \citet{xie24-MI-strategy} for strategy rule creation, client state inference, and response generation.

\subsection{Details of the User Study}
\label{app:subsec_details_user_study}

We conducted the experiment using crowdsourcing. Participants engaged in a web-based, text-based dialogue with one of the assigned systems, followed by a questionnaire to evaluate the system. Upon accessing the instruction page, participants were informed that the study's objective was to evaluate the dialogue system from the perspective of MI. They were provided with an overview of MI and its goals, and were instructed on the specific conversation topic for each experiment: concerns regarding diet for single-domain (Diet-MI), and no specific topic constraints for multi-domain (AnnoMI).
Participants were then instructed to identify a personal issue related to this topic and to engage in conversations with dialogue systems. 
Before beginning, participants were required to confirm their understanding of the instructions and study content.
The full description of the instructions for the experiment is shown in Figure \ref{app_fig_userstudy_instruct}.

After this orientation, participants accessed a chat interface as shown in Figure \ref{fig:chat_UI}. The dialogue began when participants submitted an initial message describing their concern. Each participant's utterance was reflected in the conversation history panel, followed by the system's response. Participants continued the interaction by sending subsequent utterances. They were instructed to provide at least 20 utterances (40 in total, including the system's responses) and were allowed to end the conversation at any point thereafter. Upon completion, they were asked to press a ``End Chat and Save History'' button to store the dialogue log.

Next, participants proceeded to a separate form page, where they answered 16 questions assessing their impressions of the system. They were also asked to confirm their consent for the collected data to be used for research analysis and potential public release.

We recruited participants via the crowdsourcing platform Prolific\footnote{Prolific: https://www.prolific.com/}. For each experiment, we enlisted 24 participants per condition. We imposed no restrictions on demographics such as age or gender, but required that participants' primary language be English. Participants were uniquely identified using their Prolific IDs, and no personally identifiable information (PII) was collected.
Participants took approximately 30 minutes to complete both the dialogue and the questionnaire, and were compensated £5 (approximately 6.70 USD).

Table \ref{tab:app_chat_basic_stats} presents the statistics from the user study, specifically the average number of utterances per dialogue, the average number of system sentences per dialogue, the word count per system utterance, the word count per client utterance, the response generation time per utterance, and the approximate OpenAI API inference cost per dialogue. Additionally, we include the results of the CAMI-simulator for reference.

\section{Evaluation Details}

\subsection{Human Evaluation}
\label{app:subsec_human_evaluation}
We present the 16 questionnaire items used in the user study in Table \ref{tab:app_question_list}.
Additionally, the results for all questionnaire items regarding single-domain (Diet-MI), multi-domain (AnnoMI), and Combined (Diet-MI+AnnoMI) are shown in Figures \ref{fig:enquete_all_stat_NakanoDS}, \ref{fig:enquete_all_stat_AnnoMIDS}, and \ref{fig:enquete_all_stat_bothDS}.

\subsection{Annotation}
\label{app:annotation_detail}
We describe the details of the annotation process for MITI and question categorization, which was conducted to evaluate the dialogue outcomes in Section \ref{subsec:characteristic_sys}.
Each annotation unit corresponds to a single sentence, and we allowed at most one category to be assigned per sentence.

Four annotators, all of whom had received a university-level education and demonstrated a solid understanding of the principles of MI, performed the annotation of the dialogue data.
Prior to the main annotation task, they aligned their annotation criteria using a small set of dialogue data.
Afterwards, they annotated three new samples to verify inter-annotator agreement. 

For the 11 MITI categories, the inter-annotator agreement among the four annotators was 0.60 in Fleiss's $\kappa$, indicating moderate agreement. The pairwise Cohen's $\kappa$ among the four annotators ranged from 0.52 to 0.75, showing moderate to substantial agreement.
Additionally, for the 11 question categories, the inter-annotator agreement among the four annotators was 0.60 in Fleiss's $\kappa$, indicating moderate agreement. The pairwise Cohen's $\kappa$ ranged from 0.56 to 0.73, indicating moderate to substantial agreement.
Then, the dialogue data obtained from the experiment was divided between the four annotators, and each performed the annotation separately.

Table \ref{tab:app_miti4_full} presents the MITI annotation results for single-domain (Diet-MI), multi-domain (AnnoMI), and Combined (Diet-MI+AnnoMI).
Table \ref{tab:app_q_category_def} outlines the definitions of the question categories, while Table \ref{tab:app_q_category_full} displays the corresponding annotation results for the single-domain (Diet-MI), multi-domain (AnnoMI), and Combined (Diet-MI+AnnoMI).

\section{Dialogue Examples}
We present dialogue examples collected from the user study.
An example dialogue with the MI Few-shot (MI-FS) system is shown in Table \ref{tab:sample_MI-FS}. An example dialogue with the DIIR system is shown in Table \ref{tab:sample_MI-DIIR}. An example dialogue with our proposed system (SGR-MI) is shown in Table \ref{tab:sample_ours}.

In the baseline systems, MI-FS and DIIR, there are few questions related to current problems or past efforts; the majority of the dialogue instead focuses on planning. Moreover, the system tends to provide informative advice in response to user requests, and users continue to ask for information sequentially. This results in advice-oriented dialogues that are not grounded in the user's own ideas and do not adhere to MI (Motivational Interviewing) principles.

In contrast, the proposed system (SGR-MI) emphasizes eliciting the user's thoughts by asking reflective questions, even when the user explicitly requests advice. Consequently, the frequency of advice in the dialogue is reduced, and the proportion of questions is relatively high. This shift promotes a dialogue style that encourages user reflection and contemplation. However, users did not always perceive the system as actively guiding the conversation. This may be due to the system repeating responses that did not match user requests or reflect past experiences, as illustrated by utterance ID 20-Co in Table \ref{tab:sample_ours}.

\section{Use of AI Tools}
We used the AI assistant only for polishing and proofreading the manuscript.

\lstset{
  basicstyle=\ttfamily\footnotesize,
  keywordstyle=\color{blue},
  commentstyle=\color{gray},
  stringstyle=\color{green!60!black},
  showstringspaces=false,
  breaklines=true,
  frame=single,
  captionpos=b
}

\begin{figure*}[htbp]
  \centering
  \begin{tcolorbox}[colback=white!95!gray,
                    colframe=black,
                    arc=2mm,
                    boxrule=0.5pt,
                    width=0.95\linewidth]

\begin{lstlisting}[language=Python]
from pydantic import BaseModel, Field

# --- Frame definitions ---
class GoalAndIdealFrame(BaseModel):
    frame_type: str = "goal_and_ideal"
    content: list[str] = Field(default=[], description="content of goal and ideal")

class ProblemAndTroubleFrame(BaseModel):
    frame_type: str = Field(default="problem_and_trouble")
    frame_index: int
    content: list[str] = []
    detail: list[str] = []
    harm_effect: list[str] = []
    necessity_to_improve: list[str] = []

class ExperienceFrame(BaseModel):
    frame_type: str = Field(default="experience")
    frame_index: int
    link_frame_type_and_index: list[str] = []
    content: list[str] = []
    detail: list[str] = []
    effect: list[str] = []

class ImprovementPlanFrame(BaseModel):
    frame_type: str = Field(default="improvement_plan")
    frame_index: int
    link_frame_type_and_index: list[str] = []
    content: list[str] = []
    detail: list[str] = []
    confidence_to_achive: list[str] = []


class Dialogue_State(BaseModel):
    goal_frame: GoalAndIdealFrame
    problem_and_trouble_frames: list[ProblemAndTroubleFrame] = []
    problem_and_trouble_links: list[str] = []
    experience_frames: list[ExperienceFrame] = []
    improvement_plan_frames: list[ImprovementPlanFrame] = []
\end{lstlisting}
\end{tcolorbox}

\caption{Schema definition of dialogue state in the proposed method (Pydantic format). This figure illustrates the dialogue state schema (Dialogue\_State), which structures the conversation into four frames: user's goals and ideals (GoalAndIdealFrame), problems and difficulties (ProblemAndTroubleFrame), experiences (ExperienceFrame), and improvement plans (ImprovementPlanFrame). Each frame is defined with custom attributes such as type, frame index, content, and additional properties specific to the frame type.}

\label{fig:app_DS_schema}
\end{figure*}

\begin{figure*}[htbp]
    \begin{minipage}[t]{0.5\textwidth}
    \centering
      \begin{tcolorbox}[colback=white!95!gray,
                        colframe=black,
                        arc=2mm,
                        boxrule=0.5pt,
                        width=1\linewidth]
        
        \scriptsize

        \textbf{\underline{System Message}}\\
    \#\# 1. Abstract\\
    - You are an expert in designing algorithms for information extraction. You organize and extract the content of the dialogue from the two-person counseling dialogue between the counselor and the client, while maintaining high quality and focusing on the content of the client's speech. Do not extract information that is not mentioned in the dialogue text.\\
    - Organize the dialogue state according to the following four frame types (goal\_and\_ideal, problem\_and\_trouble, experience, improvement\_plan). In addition, update the dialogue state in a process that sequentially updates the dialogue content based on partial utterances and previous dialogue state.\\
    
    \#\# 2. Explanation of attributes and attribute's values taken by the four types of frames\\
    
    goal\_and\_ideal\_frame: {\\
    frame\_type: "goal\_and\_ideal",\\
    content: [], // a general description of the client's ideals and goals\\
    }\\
    problem\_and\_trouble\_frame: {\\
    frame\_type: "problem\_and\_trouble",\\
    frame\_index, // index number of the frame\\
    content: [], // general description of the client's problem\\
    detail: [], // background and details about the client's problem\\
    harm\_effect: [], // harm or effect related to the client's problem\\
    necessity\_to\_improve: [] // describes the need to improve the client's problem\\
    }\\
    experience\_frame: {\\
    frame\_type: "experience",\\
    frame\_index, // index number of the frame\\
    link\_frame\_type\_and\_index: [] // target frame\_type and frame\_index if linked to a specific problem\_and\_trouble\\
    content: [], // general description of actions or experiences already attempted in response to a specific problem\\
    detail: [], // more detailed explanation of the actions or experiences previously attempted in response to the problem\\
    effect: [] // the effect or impact of the action or experiences that previously attempted in response to the problem\\
    }\\
    improvement\_plan\_frame: {\\
    frame\_type: "improvement\_plan",\\
    frame\_index, // index number of the frame\\
    link\_frame\_type\_and\_index: [] // Target frame\_type and frame\_index if associated with a specific problem\_and\_trouble or experience\\
    content: [], // general description of the proposed improvement for a certain problem or experience\\
    detail: [], // more detailed information about the proposed improvement for a given trouble or experience\\
    confidence\_to\_achive: [] // expresses the confidence and feasibility of the proposed improvement for a problem or experience\\
    }\\
    \#\# 3. Input/output format\\
    \#\#\# Current\_DS\\
    // Dialogue state immediately before input utterance\\
    
    \#\#\# Utterance\\
    // Past utterances and latest input utterance at the end. one utterance format: <Speaker (Counselor | Client)>: <utterance>\\
    
    \#\#\# Updated\_DS\\
    // Updated dialogue state based on the input utterance\\
    
    \textit{(The prompt is continued in the next column.)}
    
    \end{tcolorbox}
    \end{minipage}
    \hfill
    \begin{minipage}[t]{0.5\textwidth}
    \vspace{-501.5pt}
    \begin{tcolorbox}[colback=white!95!gray,
                        colframe=black,
                        arc=2mm,
                        boxrule=0.5pt,
                        width=1\linewidth]
    
    \scriptsize
    \#\# 4. Rules\\
    - Each frame type: problem\_and\_trouble, experience, improvement\_plan can be a frame with specific contents. goal\_and\_ideal\_frame must be represented by a single frame.\\
    - The granularity of the content should be as fine as possible and divided into multiple frames.\\
    - The granularity of the content should be as shown in the following example. I have a problem with drinking too much alcohol. I am very busy at work and I get stressed easily. Since “content” describes a single matter, extract "I drink too much alcohol", "I am busy at work", and "I get stressed out easily" as contents individually. These contents are considered to be independent, and each of them is considered to be a content of the problem\_and\_trouble frame.\\
    - The experience\_frame and improvement\_plan are linked to the specific problem\_and\_trouble and experience\_frame (linked frame\_type-<frame\_index>).\\
    - If frame of problem\_and\_trouble are related to each other based on the content of the dialogue, the two related problem\_and\_trouble frames should be listed in the problem\_and\_trouble\_links variable. For example, if problem\_and\_trouble-1 and problem\_and\_trouble-2 are related, describe them as problem\_and\_trouble\_links: ["problem\_and\_trouble-1\&2"].\\
    - The attribute's values for each frame are bulleted and can take multiple attribute's values. If it is not mentioned in the dialogue, no frame is created, and the attribute's value of the frame is assumed to be empty ([]) if there is no corresponding content.\\
    - If there is no dialogue state to be updated from the utterance, output the contents of Current\_DS as is as Updated\_DS.\\
    
    {-}{-}{-} Examples {-}{-}{-}\\
    \#\#\# Current\_DS\\
    <Current\_DS>\\
    
    \#\#\#  Utterance\\
    Counselor: \\
    Client: \\
    
    \#\#\# Updated\_DS\\
    <Updated\_DS>\\
    
    {-}{-}{-}\\
    ... \textit{(9 other examples follow in the same manner)}\\
    
    {-}{-}{-} Example End {-}{-}{-}\\
        \tcblower

        \scriptsize
        \textbf{\underline{User Message}}\\
    \#\#\# Current\_DS\\
    <Current\_DS>\\
    
    \#\#\# Utterance\\
    <Utterance (Use the most recent six utterances from the dialogue history between a counselor and a client, formatted as: <Speaker (Counselor | Client)>: <utterance>)>\\
    
    \#\#\# Updated\_DS\\
    \end{tcolorbox}
    
    \end{minipage}
    \caption{Prompt for dialogue state generation. Ten examples of dialogue state updates (each consisting of a pair of pre-update dialogue state and dialogue history as input, and post-update dialogue state as output) are provided. The current dialogue context is given as a User Message. During generation, the input is passed to the LLM via function calling, using the schema defined in Figure \ref{fig:app_DS_schema}, to generate the dialogue state.}
    
    \label{fig:app_extract_DS_prompt}
    \end{figure*}

\lstset{
  basicstyle=\ttfamily\footnotesize,
  keywordstyle=\color{blue},
  commentstyle=\color{gray},
  stringstyle=\color{green!60!black},
  showstringspaces=false,
  breaklines=true,
  frame=single,
  captionpos=b
}

\begin{figure*}[htbp]
\centering

\begin{minipage}[t]{0.95\linewidth}
\begin{tcolorbox}[colback=white!95!gray,
                    colframe=black,
                    arc=2mm,
                    boxrule=0.5pt,
                    width=1\linewidth]
\footnotesize

\textbf{\underline{System Message}}\\
Based on the counselor's response and the preceding dialogue history and dialogue state, output the dialogue strategy for the counselor response, as described below.\\
\#\#\# Input information\\
- Dialogue History (History): A set of recent utterances between the counselor and the client.\\
- Dialogue State (Dialogue\_State): Structured representation of the content of the dialogue up to the preceding utterance.\\
- Counselor Response (Counselor\_response): The response provided by the counselor.\\

\#\#\# Output information\\
Dialogue Strategy for Counselor Response\\
Dialogue\_Strategy: \{
\vspace{-1\baselineskip}
\begin{description}
\setlength{\parskip}{0pt}
\item[] focuses: From Dialogue\_State, extract frames of the dialog state that the counselor's response is focused on. The values are references to a specific frame. Each value consists of frame\_type, frame\_index, and frame\_attribute. For example, if the response pertains to the entire content of a particular frame, the format will be <frame\_type>, <frame\_index>, frame\_attribute="all". If it targets a specific attribute of an instance, the format will be <frame\_type>, <frame\_index>, frame\_attribute="name of the target attribute".
\item[] seek\_frame\_type: If the counselor's response aims to elicit a specific frame\_type from the user within the Dialogue\_State, specify the frame\_type. Possible values include: goal\_and\_ideal, problem\_and\_trouble, experience, improvement\_plan.
\item[] seek\_attribute: If the counselor's response aims to elicit a specific attribute from the seek\_frame\_type, specify the attribute name. Possible values include: detail, harm\_effect, necessity\_to\_improve, effect, confidence\_to\_achieve.
\end{description}
\vspace{-1\baselineskip}

\}\\

\tcblower
\footnotesize
\textbf{\underline{User Message}}\\
\#\#\# History\\
<History>\\

\#\#\# Dialogue\_State\\
<Dialogue\_State>\\

\#\#\# Counselor\_response\\
<Counselor\_response>

\end{tcolorbox}
\end{minipage}

\vspace{1\baselineskip}

\begin{minipage}{0.8\linewidth}
\begin{lstlisting}[language=Python]
from pydantic import BaseModel, Field

class Focus(BaseModel):
    frame_type: str
    frame_index: Optional[int]
    frame_attribute: Optional[Union[str, Literal['all']]]

class DialogueStrategyWithoutIntent(BaseModel):
    focuses: list[Focus]
    seek_frame_type: Optional[str]
    seek_attribute: Optional[str]

\end{lstlisting}
\end{minipage}

\caption{Prompt for generating pseudo dialogue strategy samples and the schema of output dialogue strategy (in Pydantic format).}

\label{fig:app_policy_gen_pseudo}
\end{figure*}

\begin{figure*}[htbp]
\centering

\begin{minipage}[t]{0.95\linewidth}
\begin{tcolorbox}[colback=white!95!gray,
                    colframe=black,
                    arc=2mm,
                    boxrule=0.5pt,
                    width=1\linewidth]
\scriptsize

\textbf{\underline{System Message}}\\
You are a counselor conducting motivational interviews related to concerns about diet and eating habits. Determine the intention and content of the counselor's next utterance. Below is a description of the input and output formats.
Examples of similar dialogue situations are presented under ``Examples'' based on the subsequent input, determine the output.

\#\#\# Input information\\
- Dialogue History (History): A set of recent utterances between the counselor and the client.\\
- Dialogue State (Dialogue\_State): Structured representation of the content of the dialogue up to the preceding utterance.\\

\#\#\# Output information\\
Dialogue\_Strategy: \{
\vspace{-1\baselineskip} 
\begin{description}
\setlength{\parskip}{0pt}
\item[] intent: The intent refers to the next utterance intention of the Counselor and can take one of the following values: Question, Affirmation, Reflection, Summarization, or Other. Each is defined as follows. Question refers to utterances that ask the other person a question. Affirmation includes utterances that acknowledge or praise the other person. Reflection refers to utterances that repeat, rephrase, or reflect back what the other person said, possibly adding concrete metaphors or imagined situations. Summarization represents utterances that summarize or paraphrase two or more previous statements. Other refers to utterances that do not fall into any of the above categories.
\item[] focuses: From Dialogue\_State, extract frames of the dialog state that the counselor's response is focused on. The values are references to a specific frame. Each value consists of frame\_type, frame\_index, and frame\_attribute. For example, if the response pertains to the entire content of a particular frame, the format will be <frame\_type>, <frame\_index>, frame\_attribute="all". If it targets a specific attribute of an instance, the format will be <frame\_type>, <frame\_index>, frame\_attribute="name of the target attribute".
\item[] seek\_frame\_type: If the counselor's response aims to elicit a specific frame\_type from the user within the Dialogue\_State, specify the frame\_type. Possible values include: goal\_and\_ideal, problem\_and\_trouble, experience, improvement\_plan.
\item[] seek\_attribute: If the counselor's response aims to elicit a specific attribute from the seek\_frame\_type, specify the attribute name. Possible values include: detail, harm\_effect, necessity\_to\_improve, effect, confidence\_to\_achieve.
\end{description}
\vspace{-1\baselineskip}
\}

\tcblower
\scriptsize

\textbf{\underline{User Message}}\\
{-}{-}{-} Examples {-}{-}{-}\\ \\
\textit{}
\#\#\# History\\
<Example1's History>\\

\#\#\# Dialogue\_State\\
<Example1's Dialogue\_State>\\

\#\#\# Dialogue\_Strategy\\
<Example1's Dialogue\_Strategy>\\

{-}{-}{-}\\
... \textit{(4 other examples follow in the same manner)}\\
{-}{-}{-} Examples End {-}{-}{-}\\ \\
\#\#\# History\\
<History>\\

\#\#\# Dialogue\_State\\
<Dialogue\_State>\\

\#\#\# Dialogue\_Strategy

\end{tcolorbox}
\end{minipage}

\begin{minipage}{0.8\linewidth}
\begin{lstlisting}[language=Python]
from pydantic import BaseModel, Field

class Focus(BaseModel):
    frame_type: str
    frame_index: Optional[int]
    frame_attribute: Optional[Union[str, Literal['all']]]

class DialogueStrategy(BaseModel):
    intent: str
    focuses: list[Focus]
    seek_frame_type: Optional[str]
    seek_attribute: Optional[str]

\end{lstlisting}
\end{minipage}

\caption{Prompt for generating dialogue strategies used in system responses and the schema of the output dialogue strategy (in Pydantic format). Unlike the pseudo dialogue strategy generation (Figure \ref{fig:app_policy_gen_pseudo}), this schema includes an intent attribute. To refer dialogue strategies used in situations similar to the current dialogue context, we retrieve samples from a pool of pseudo dialogue strategies that are closest to the current context and use them as few-shot demonstrations (see Examples section).}

\label{app_policy_gen}
\end{figure*}

\lstset{
  basicstyle=\ttfamily\footnotesize,
  keywordstyle=\color{blue},
  commentstyle=\color{gray},
  stringstyle=\color{green!60!black},
  showstringspaces=false,
  breaklines=true,
  frame=single,
  captionpos=b
}

\begin{figure*}[htbp]
\centering

\begin{minipage}[t]{0.95\linewidth}
\begin{tcolorbox}[colback=white!95!gray,
                    colframe=black,
                    arc=2mm,
                    boxrule=0.5pt,
                    width=1\linewidth]
\scriptsize

\textbf{\underline{System Message}}\\
You are a counselor conducting motivational interviewing for clients with concerns about their diet and eating habits. Based on the information provided below, generate the counselor's next response. The response should be no more than two sentences. Each item is described as follows:\\
- Dialogue History (History): A set of recent utterances between the counselor and the client.\\
- Dialogue State (Dialogue\_State): Structured representation of the content of the dialogue up to the preceding utterance.\\
- Next Utterance Intent (Intent): The communicative intent of the next response to generate. Possible values include: Question, Affirmation, Reflection, Summarization, Other.\\
- Focuses for the Next Utterance (Focuses): The frames of the dialogue state to be referred to in the responses to be generated. \\
- Seek Frame Type (Seek\_frame\_type): The frame type to elicit from the client in the next utterance.\\
- Seek Attribute (Seek\_attribute): The attribute of the frame that the next utterance will attempt to elicit from the client.\\

\#\#\# Explanation of attributes and attribute's values taken by the four types of frames\\
goal\_and\_ideal\_frame: \\
- frame\_type: "goal\_and\_ideal",\\
- content: [], // a general description of the client's ideals and goals\\
\\
problem\_and\_trouble\_frame: \\
- frame\_type: "problem\_and\_trouble",\\
- frame\_index, // index number of the frame\\
- content: [], // general description of the client's problem\\
- detail: [], // background and details about the client's problem\\
- harm\_effect: [], // harm or effect related to the client's problem\\
- necessity\_to\_improve: [] // describes the need to improve the client's problem\\
\\
experience\_frame: \\
- frame\_type: "experience",\\
- frame\_index, // index number of the frame\\
- link\_frame\_type\_and\_index: [] // target frame\_type and instance index if linked to a specific problem\_and\_trouble\\
- content: [], // general description of actions or experiences already attempted in response to a specific problem\\
- detail: [], // more detailed explanation of the actions or experiences previously attempted in response to the problem\\
- effect: [] // the effect or impact of the action or experiences that previously attempted in response to the problem\\
\\
improvement\_plan\_frame: \\
- frame\_type: "improvement\_plan",\\
- frame\_index, // index number of the frame\\
- link\_frame\_type\_and\_index: [] // Target frame\_type and instance index if associated with a specific problem\_and\_trouble or experience\\
- content: [], // general description of the proposed improvement for a certain problem or experience\\
- detail: [], // more detailed information about the proposed improvement for a given trouble or experience\\
- confidence\_to\_achive: [] // expresses the confidence and feasibility of the proposed improvement for a problem or experience\\
\\
\#\#\# Caution\\
When the Intent is Reflection or Summarization, do not ask for new information; instead, respond with content that reflects or summarizes.
If Seek\_attribute = content, it means the client is requesting a new Seek\_frame\_type information. In such cases, ignore the details in Focuses and respond by prompting a new frame of topic (e.g., for Seek\_frame\_type: problem\_and\_trouble, Seek\_attribute = content, you should ask something like ``Is there anything else troubling you?'' without using Focuses).\\
Refer to the dialogue history as well, and avoid repeating similar responses already made by the Counselor. If a similar response is inevitable, consider changing the Intent category to maintain novelty.\\

\#\#\# Flow of the dialogue\\
Initially, focus on identifying the client's problems and what efforts they have already made. Then, focus on a particular issue and collaboratively discuss improvement ideas. Finally, evaluate the feasibility of the proposed solutions with the client.

\tcblower
\scriptsize
\textbf{\underline{User Message}}\\
\#\#\# History\\
<History>\\

\#\#\# Dialogue\_State\\
<Dialogue\_State>\\

\#\#\# Intent\\
<intent>\\

\#\#\# Focuses\\
<focuses>\\

\#\#\# Seek\_frame\_type\\
<seek\_frame\_type>\\

\#\#\# Seek\_attribute\\
<seek\_attribute>\\

\#\#\# Response

\end{tcolorbox}
\end{minipage}

\caption{Prompt for generating a system response.}

\label{app_RG_prompt}
\end{figure*}

\begin{figure*}[htbp]
  \centering
  \begin{tcolorbox}[colback=white!95!gray,
                    colframe=black,
                    arc=2mm,
                    boxrule=0.5pt,
                    width=0.95\linewidth]

    \footnotesize
    \textbf{\underline{System Message}}\\
    The following are examples of how a counselor might respond to a client based on the context of their previous interactions. Learn from these examples and generate a counselor's response to the final example. The response should be no more than two sentences.\\ \\
    \# Example 1\\
    \#\# Context\\
    Client: ...\\
    Counselor: ...\\
    ...\\
    Client: ...\\
    \#\# Response\\
    Counselor: ...\\ \\
    \# Example 2\\
    ...\\ \\
    \# Example 3\\
    ...\\ \\
    \# Example 4\\
    ...\\ \\
    \# Example 5\\
    ...
    
    \tcblower
    \footnotesize
    \textbf{\underline{User Message}}\\
    \# Example 6\\
    \#\# Context\\
    Clinet: ...\\
    Counselor: ...\\
    ...\\
    Client: ...\\
    \#\# Response\\
    Counselor: 
  \end{tcolorbox}
    \caption{Prompt used in MI Few-shot (MI-FS). Based on the prompt settings used in \citet{wu-fs-prompt-ack23}, we used 15 utterances between the client and counselor in the \#\# Context section, and inserted the subsequent counselor's response in the ``Counselor:'' part under \#\# Response section. We used five such examples, each extracted from real conversations conducted by professional counselors. In Example 6 of the User Message, we provided the model with 15 previous utterances from the ongoing MI conversation as context and had the LLM generate the subsequent counselor response.}
    
\label{fig:app_MI_fs_prompt}
\end{figure*}

\clearpage

\lstset{
  basicstyle=\ttfamily\footnotesize,
  keywordstyle=\color{blue},
  commentstyle=\color{gray},
  stringstyle=\color{green!60!black},
  showstringspaces=false,
  breaklines=true,
  frame=single,
  captionpos=b
}

\begin{figure*}[htbp]
\centering

\begin{minipage}[t]{0.95\linewidth}
\begin{tcolorbox}[colback=white!95!gray,
                    colframe=black,
                    arc=2mm,
                    boxrule=0.5pt,
                    width=1\linewidth]

\footnotesize

In this experiment, we will ask you to interact with the system and evaluate the content of the dialogue from a counseling perspective called \textbf{Motivational Interviewing}. The purpose of this experiment is to evaluate whether the dialogue was conducted in accordance with the Motivational Interviewing method.\\ \\

Motivational interviewing is a counseling method that helps people facilitate behavioral change related to problems they want to address in their daily lives, particularly for those who struggle to change their behaviors despite knowing they should. For example, "I know that drinking too much alcohol is not good for my health, but I can't help but drink too much," is a topic that is addressed in Motivational Interviewing.\\ 

Motivational interviewing uses a collaborative style of conversation to review current mental health issues, motivate the client to change the behaviors on his/her own initiative, and strengthen their commitment.\\

\textbf{The topic in this experiment is not specified.} Please have a conversation with the AI about things in your daily life that you would like to change or improve. You are asked to develop a plan to address the problem through dialogue with the system. The system will support your behavior change by asking questions and offering thought-provoking responses to your concerns.  \\

\textbf{<Example of topics>}\\
• Unable to control the amount of food I eat\\
• Trying to cut down on smoking\\
• I tend to drink too much alcohol on a daily basis\\
*You are welcome to discuss any other concerns that you would like to improve.\\

\textbf{<Precautions>}\\
- There are no right or wrong answers in the dialogue with the system. Please proceed with the conversation naturally at your own pace.\\
- This system uses AI technology to generate responses. There is a possibility that the system's response sentences, advice, etc. may contain lies, errors, or inappropriate responses. Please understand that such output is possible before working on this experiment.\\
- Please use a PC browser for the working environment. The system uses JavaScript. The recommended browser is Google Chrome, and we have verified its operation with Google Chrome.\\

\end{tcolorbox}
\end{minipage}

\caption{Instruction of the experiment before the start of the dialogue in the user study. This figure displays the instructions for the multi-domain (AnnoMI) sessions (no specific topic constraint). In the single-domain (Diet-MI) sessions, the topic was restricted to ``diet'' and the ``Example of topics'' section listed diet-related concerns.}

\label{app_fig_userstudy_instruct}
\end{figure*}

\begin{figure*}[h]
  \centering
  \includegraphics[keepaspectratio,scale=0.6]{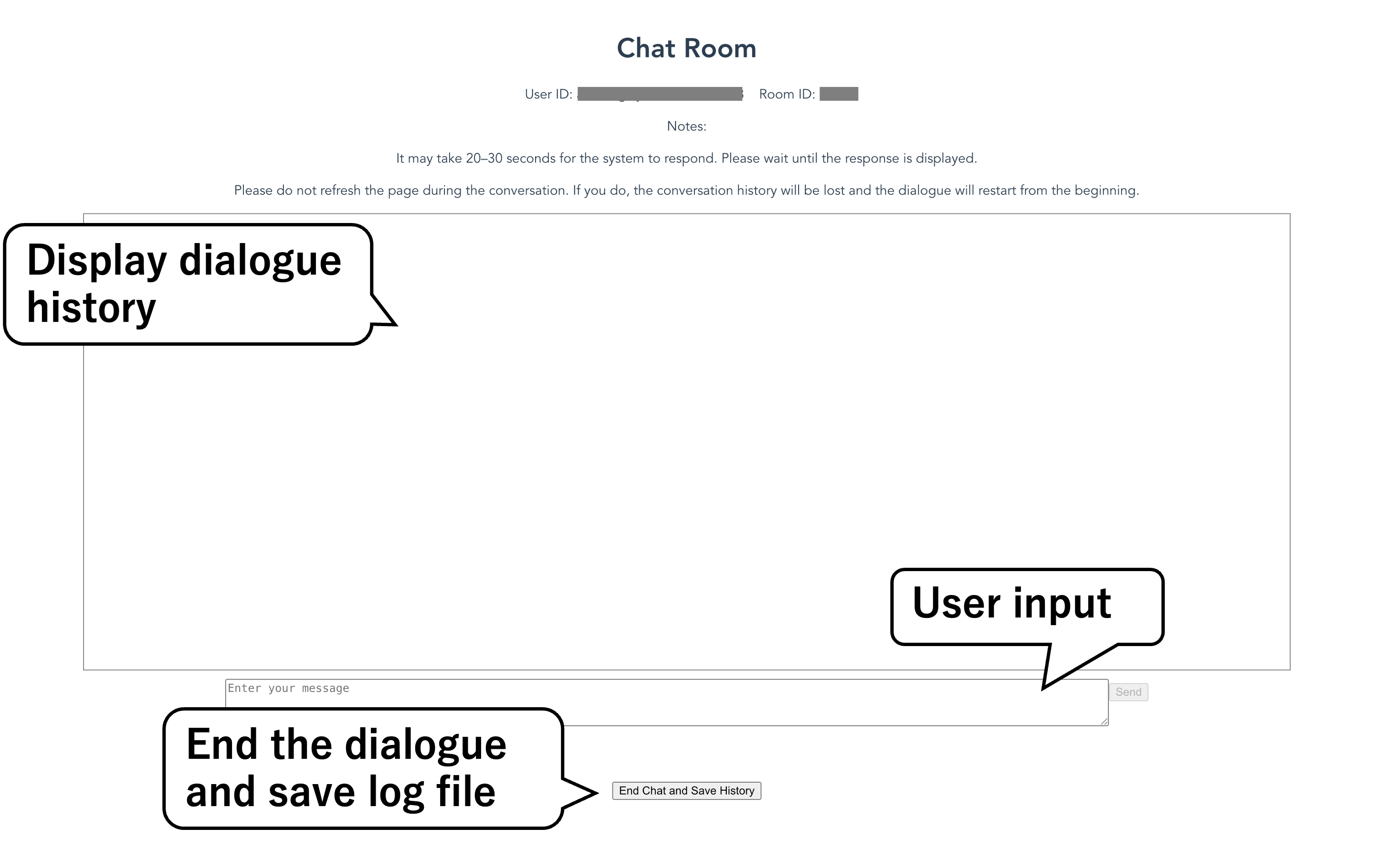}
  
  \caption{Chat page}
  \vspace*{-0.5cm} 
  \label{fig:chat_UI}
\end{figure*}

\begin{table*}[]
\resizebox{1\linewidth}{!}{%

\begin{tabular}{p{45mm}p{18mm}p{18mm}p{18mm}|p{25mm}}
\toprule
    &  MI-FS  & DIIR  & SGR-MI & CAMI-simulator \\
\midrule

Utterances (user+system) per dialogue & 43.0 (8.9) & 41.5 (2.4) & 41.7 (2.8) & 27.5 (12.2) \\
\addlinespace[3mm]
System sentences per dialogue         & 45.9 (10.2) & 36.0 (3.0)  & 22.6 (2.4) & 47.9 (22.1)\\
Number of words in system utterance   & 38.7 (3.6) & 33.4 (2.5) & 23.7 (3.6) & 41.3 (0.7) \\
Number of words in client utterance   & 15.4 (5.9) & 14.8 (7.0) & 16.1 (9.0) & 44.6 (1.7) \\
System response generation time per utterance (sec) & 1.8 (0.9) & 5.0 (0.7) & 8.8 (5.5) & 28.7 (4.5) \\
Approximate system inference cost per dialogue (USD) & 0.07 & 0.19 & 0.28 & 0.17 \par(include client) \\
\bottomrule

\end{tabular}
}
\caption{Basic dialogue statistics obtained from each system. The numbers in parentheses indicate the standard deviation.}
\label{tab:app_chat_basic_stats}
\end{table*}

\begin{table*}[h]
\centering
\begin{tabular}{ll} \toprule
ID & Question \\ \midrule
Q1 & The system responded based on what you said and discussed.\\
Q2 & The system remembers what you have said and discussed. \\
Q3 & The system repeated questions that have already been asked or provide responses \\
   & with the same content. \\
Q4 & The system steered the conversation in one direction and rushed to a conclusion.\\
Q5 & The system asked questions that made you think deeply about the current issue.\\
Q6 & You talked with the system about how you have been dealing with the current problems and \\ 
   & about your past efforts and experiences. \\
Q7 & The system proposed ideas or solutions for improving the issue.\\
Q8 & The system frequently asked for your opinions and thoughts.\\
Q9 & The system seemed to encourage you to think for yourself rather than just giving answers.\\
Q10 & When you asked questions, the system often responded with suggestions for improvement.\\
Q11 & The system frequently offered proactive advice or suggestions.\\
Q12 & You often asked the system for ideas.\\
Q13 & The system often asked you for ideas.\\
Q14 & You were able to create an improvement plan based on your own ideas.\\
Q15 & The system asked about your confidence in the plan for improvement.\\
Q16 & You felt motivated to make changes through your interaction with the system. \\ \bottomrule
\end{tabular}

\caption{List of subjective evaluation questions}
\label{tab:app_question_list}
\end{table*}

\begin{figure*}[h]
  \centering
  \includegraphics[keepaspectratio,scale=0.8]{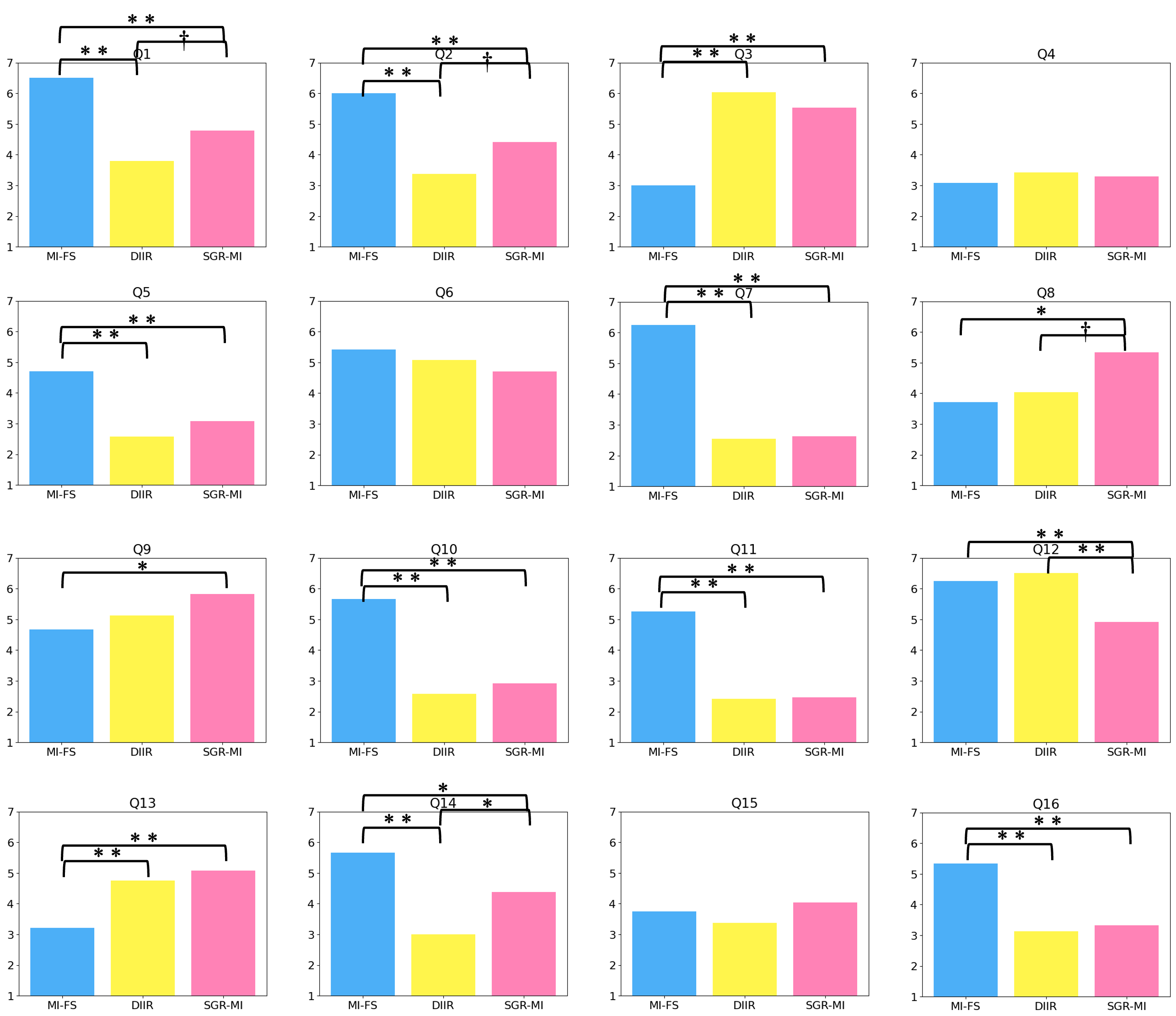}
  \caption{Subjective evaluation of 16 questions in single-domain (Diet-MI). We computed the p-value using a Tukey's HSD test. ($\dag: p<.1,\:{*}:p<.05,\:{*}{*}:p<.01$)}
  \vspace*{-0.5cm}              
  \label{fig:enquete_all_stat_NakanoDS}
\end{figure*}

\begin{figure*}[h]
  \centering
  \includegraphics[keepaspectratio,scale=0.8]{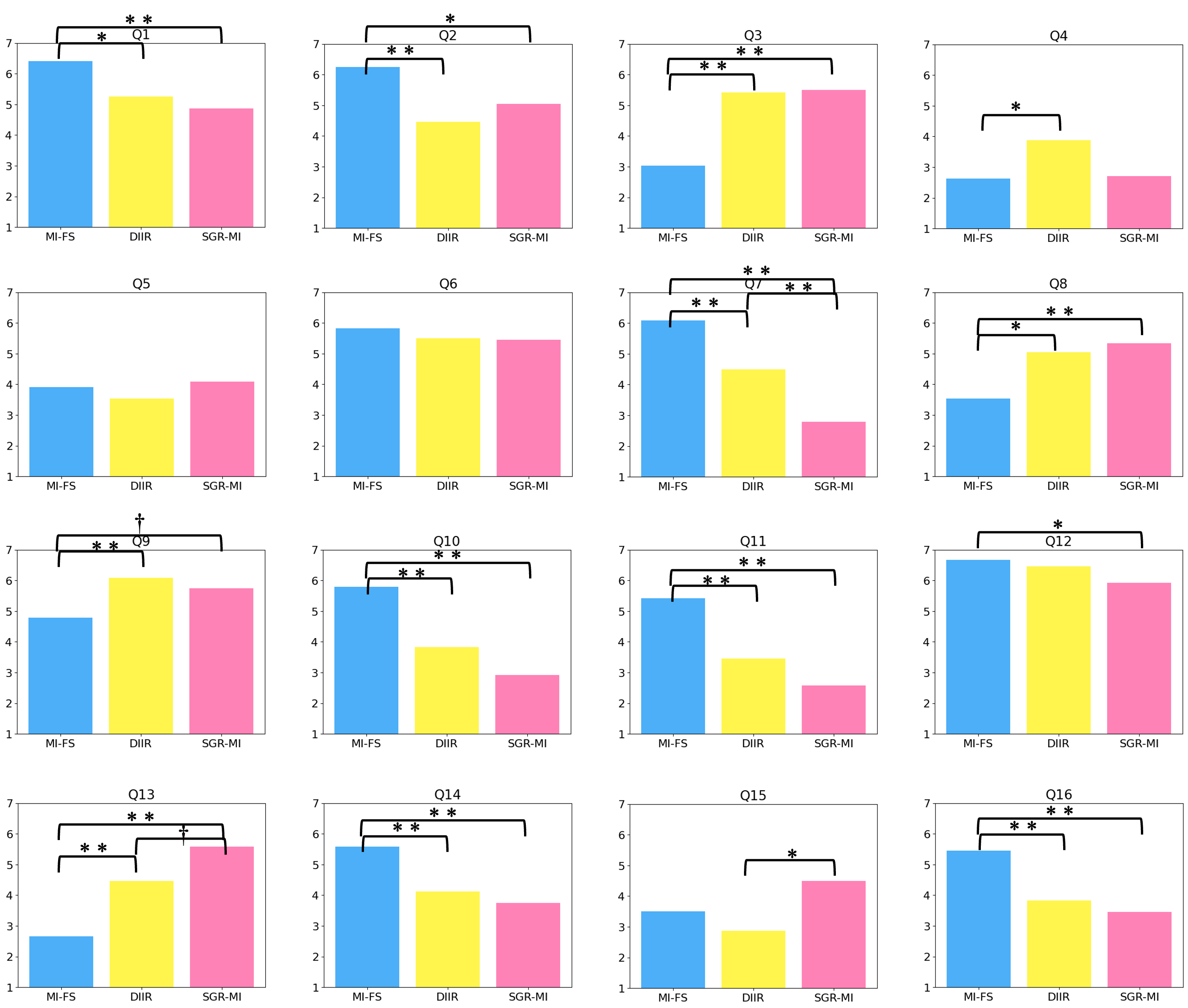}

  \caption{Subjective evaluation of 16 questions in multi-domain (AnnoMI). We computed the p-value using a Tukey's HSD test. ($\dag: p<.1,\:{*}:p<.05,\:{*}{*}:p<.01$)}
  \vspace*{-0.5cm}              
  \label{fig:enquete_all_stat_AnnoMIDS}
\end{figure*}

\begin{figure*}[h]
  \centering
  \includegraphics[keepaspectratio,scale=0.8]{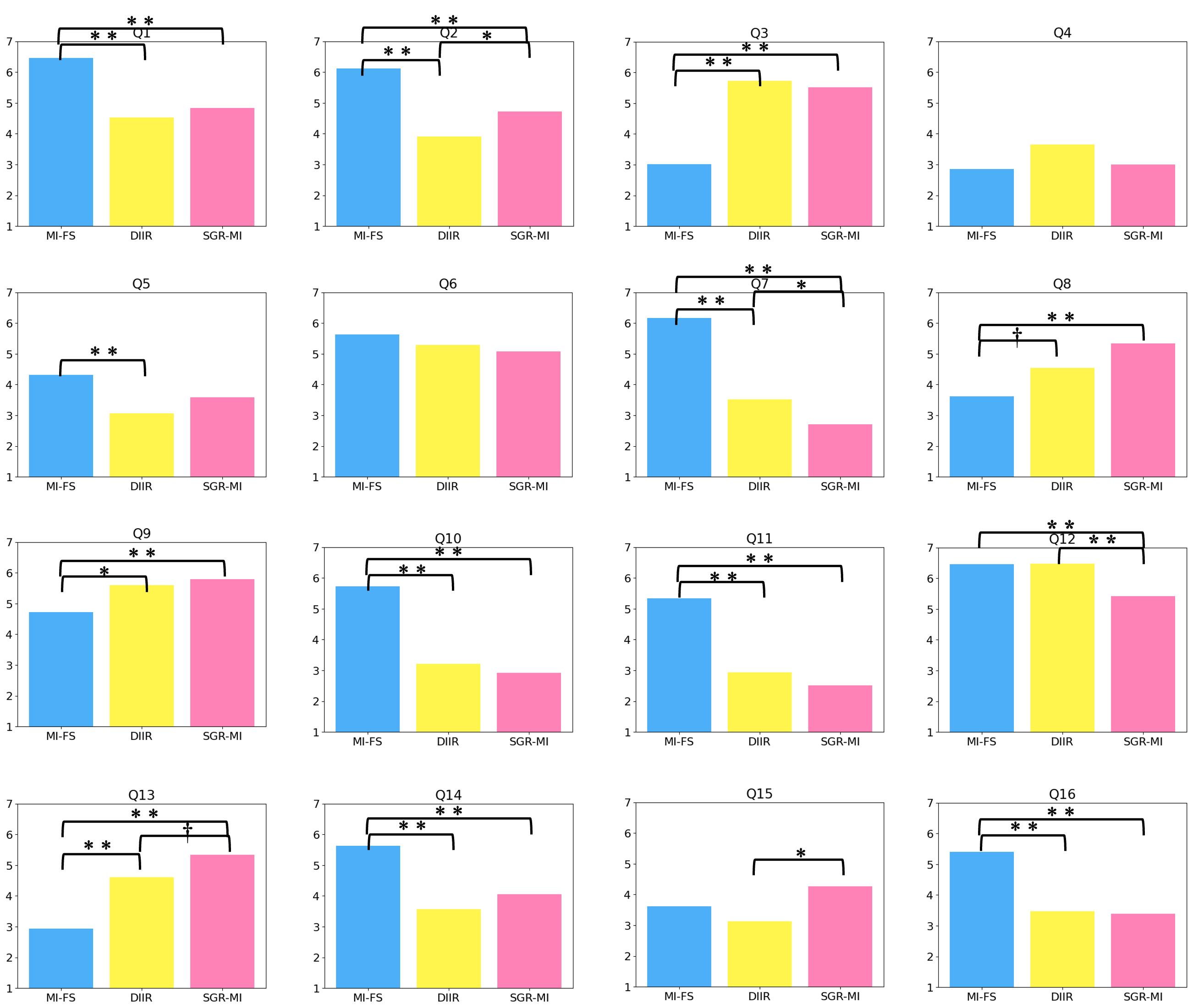}

  \caption{Subjective evaluation results for the 16 questions in Combined (Diet-MI+AnnoMI).
 We computed the p-value using a Tukey's HSD test. ($\dag: p<.1,\:{*}:p<.05,\:{*}{*}:p<.01$)}
  \vspace*{-0.5cm}              
  \label{fig:enquete_all_stat_bothDS}
\end{figure*}

\clearpage

\begin{table*}[]
\resizebox{0.95\linewidth}{!}{%

\begin{tabular}{lrrrrrrrrrrrr}
\toprule
\colorbox{gray!30}{\shortstack{Single-domain\\(Diet-MI)}}  & gi    & q     & sr   & cr   & af & ea & c   & seek  & per   & pwp   & n/a   & \shortstack{Total\\sentences} \\ 
\midrule
\multirow{2}{*}{MI-FS}  & 0.146 & 0.075	& 0.040 & 0.030 & 0.022  & 0.001  & 0   & 0.005 & 0.097 & 0.424 & 0.16  & \\
                        & (153)	& (79)	& (42)	& (32)	& (23)	 & (1)	  & (0)	& (5)	& (102)	& (445)	& (168) & (1050) \\
\addlinespace[2mm]   
\multirow{2}{*}{DIIR}   & 0.012 & 0.179 & 0.218 & 0.144  & 0.126 & 0      & 0   & 0.006 & 0.059 & 0.172  & 0.085 & \\
                        & (10)  & (150) & (182) & (120)  & (105) & (0)    & (0) & (5)   & (49)  & (144)  & (71)  & (836) \\
\addlinespace[2mm]   
\multirow{2}{*}{SGR-MI}   & 0     & 0.433 & 0.276 & 0.147  & 0.066 & 0.004  & 0   & 0     & 0     & 0.044  & 0.029 & \\
                        & (0)   & (235) & (150) & (80)   & (36)  & (2)    & (0) & (0)   & (0)   & (24)   & (16)  & (543) \\
\addlinespace[2mm]   
\multirow{2}{*}{\shortstack{SGR-MI w/o\\Strategy}}   & 0.065  & 0.323 & 0.055 & 0.034  & 0.034 & 0      & 0   & 0.040  & 0.171   & 0.207  & 0.071 & \\
                                                     & (48)   & (237) & (40)  & (25)   & (25)  & (0)    & (0) & (29)   & (125)   & (152)   & (52)  & (733) \\
\bottomrule
\addlinespace[5mm]   

\toprule
\colorbox{gray!30}{\shortstack{Multi-domain\\(AnnoMI)}}  & gi    & q      & sr   & cr   & af & ea & c  & seek  & per   & pwp     & n/a   & \shortstack{Total\\sentences} \\ 
\midrule
\multirow{2}{*}{MI-FS}  & 0.066 & 0.040  & 0.021 & 0.030  & 0.041 & 0.001  & 0  & 0.005 & 0.099 & 0.467  & 0.230 & \\
                        & (76)  &(46)    & (24)  & (35)   & (47)  & (1)    &(0) & (6)   & (114) & (539)  & (265) & (1153) \\
\addlinespace[2mm]   
\multirow{2}{*}{DIIR}   & 0.018 & 0.129  & 0.099 & 0.086  & 0.127 & 0.001  & 0  & 0.010 & 0.058 & 0.288  & 0.184 & \\
                        & (16)  & (115)  & (88)  & (77)   & (113) & (1)    & (0)& (9)   & (52)  & (257)  & (164) & (892) \\
\addlinespace[2mm]   
\multirow{2}{*}{SGR-MI}   & 0.002 & 0.459 & 0.265 & 0.151   & 0.028 & 0      & 0  & 0.002 & 0     & 0.052  & 0.042 & \\
                        & (1)	& (249)	& (144)	& (82)	  & (15)  & (0)    & (0)& (1)	& (0)	& (28)   & (23)	& (543) \\
\addlinespace[2mm]   
\multirow{2}{*}{\shortstack{SGR-MI w/o\\Strategy}}   & 0.028 & 0.406    & 0.044 & 0.062   & 0.062 & 0      & 0  & 0.015 & 0.103 & 0.167   & 0.113   & \\
                                                     & (20)	 & (289)	& (31)	& (44)	  & (44)  & (0)    & (0)& (11)	& (73)	& (119)   & (80)	& (711) \\
\bottomrule
\addlinespace[5mm]   

\toprule
\colorbox{gray!30}{\shortstack{Combined\\(Diet-MI+AnnoMI)}}  & gi    & q      & sr   & cr   & af & ea & c  & seek  & per   & pwp     & n/a   & \shortstack{Total\\sentences} \\ 
\midrule
\multirow{2}{*}{MI-FS}  & 0.104 & 0.057 & 0.030 & 0.030   & 0.032 & 0.001  & 0  & 0.005 & 0.098 & 0.447  & 0.197 & \\
                        & (229) & (125) & (66)  & (67)     & (70) & (2)    & (0)& (11)  & (216) & (984)  & (433) & (2203) \\
\addlinespace[2mm]   
\multirow{2}{*}{DIIR}   & 0.015 & 0.153 & 0.156 & 0.114   & 0.126 & 0.001  & 0  & 0.008 & 0.058 & 0.232  & 0.136 & \\
                        & (26)	& (265)	& (270)	& (197)	  & (218) & (1)	   & (0)& (14)	& (101)	& (401)	& (235)	& (1728) \\ 
\addlinespace[2mm]   
\multirow{2}{*}{SGR-MI}   & 0.001 & 0.446 & 0.271 & 0.149 & 0.047 & 0.002 & 0 & 0.001 & 0 & 0.048 & 0.036 & \\
                        & (1)	& (484)	& (294)	& (162)	& (51)	& (2)	& (0)	& (1)	& (0)	& (52)	& (39)	& (1086) \\
\addlinespace[2mm]   
\multirow{2}{*}{\shortstack{SGR-MI w/o\\Strategy}}   & 0.047 & 0.364    & 0.049 & 0.048 & 0.048 & 0    & 0      & 0.028 & 0.137 & 0.188 & 0.091 & \\
                                                     & (68)	 & (526)	& (71)	& (69)	& (69)	& (0)  & (0)	& (40)	& (198)	& (271)	& (132)	& (1444) \\
\bottomrule
\end{tabular}
}

\caption{Proportion of counselor behavior categories based on MITI \citep{moyers_16_MITI4} within system utterance sentences. The numbers in parentheses indicate raw counts. The behavior tags are as follows: gi (Giving Information), q (Questions), sr (Simple Reflection), cr (Complex Reflection), af (Affirm), ea (Emphasize Autonomy), c (Confront), seek (Seek Collaboration), per (Persuasion), pwp (Persuade with Permission), and n/a (not applicable)}

\label{tab:app_miti4_full}
\end{table*}

\begin{table*}[]
\resizebox{0.95\linewidth}{!}{%

\begin{tabular}{p{15mm}p{35mm}p{110mm}}
    \toprule
        & \multicolumn{2}{l}{\textit{Evoking questions}}   \\
        & Desire (ev-d) & e.g. How would you like things to change? \\ 
        & Ability (ev-a) &  e.g. How confident are you that you can actually do it?\\
        & Reasons (ev-r) &   e.g. Why do you want more exercise?  \\
        & Need (ev-n) & e.g. How important is it to you? \\
\addlinespace[3mm]   
        & Other (ev) & Questions that do not fit into the above categories.\\ 
\addlinespace[5mm]   
\midrule
       & \multicolumn{2}{l}{\textit{Planning questions}}   \\
       & Activation (pl-a) &  e.g. Are you willing to try it?\\
       & Commitment (pl-c) &  e.g. Do you want to do it?, Is that what you are aiming to do? \\
       & More specific (pl-d) & e.g. How do you prepare and get ready? \\
       & Setting a date (pl-t) & e.g. When is it likely to be ready?\\
       & Preparing (pl-p) & e.g. What is the first step? What do you need to take with you?  \\
\addlinespace[3mm]   
      & Other (pl) & Questions that do not fit into the above categories.\\ 
       
    \bottomrule
\end{tabular}

}
\caption{Definition of question categories and examples of utterances}
\label{tab:app_q_category_def}

\end{table*}
\begin{table*}[]
\resizebox{1\linewidth}{!}{%

\begin{tabular}{lrrrrrrr|rrrrrrrr|r} 
    \toprule
    & \multicolumn{7}{l}{Process: Evoking}                                           & \multicolumn{8}{l}{Process: Planning}         & \shortstack{Total\\sentences}  \\ \cline{2-17}
                    &        & \multicolumn{4}{c}{Evoking questions} & \multicolumn{1}{|r}{}  &        &     & \multicolumn{5}{c}{Planning questions}  & \multicolumn{1}{|r}{} &      & \\ \cline{3-6} \cline{10-14}
                       & ev     & ev-d  & ev-a      & ev-r  & ev-n   & \multicolumn{1}{|r}{ev-*}  & ev*   & pl    & pl-a  & pl-c & pl-d  & pl-t  & pl-p  & \multicolumn{1}{|r}{pl-*}  & pl*   &   \\ \midrule
\multicolumn{6}{l}{\colorbox{gray!30}{Single-domain (Diet-MI)} }                & \multicolumn{1}{|r}{}      &       &       &       &      &       &       &       & \multicolumn{1}{|r}{}      &       &       \\
\multirow{2}{*}{MI-FS} &  0.055 & 0.005 & 0.002     & 0.002 & 0.001  & \multicolumn{1}{|r}{0.010} & 0.065 & 0.006 & 0.003 & 0    & 0.006 & 0 & 0     & \multicolumn{1}{|r}{0.009} & 0.014 &       \\
                       &   (58) &  (5)  &  (2)      &  (2)  & (1)    & \multicolumn{1}{|r}{(10)}   &  (68) & (6)  &  (3)  &  (0) & (6)   & (0)   & (0)   & \multicolumn{1}{|r}{(9)}   &  (15) & (1050)\\
\multirow{2}{*}{DIIR}  &  0.104 & 0.005 & 0.005     & 0.010 &  0.002 & \multicolumn{1}{|r}{0.022} & 0.126 & 0.031 & 0.001 & 0    &  0.007& 0.013 & 0.006 & \multicolumn{1}{|r}{0.028} &  0.059&       \\
                       &   (87) &  (4)  &  (4)      &  (8)  & (2)    & \multicolumn{1}{|r}{(18)}  &  (105) & (26)  &  (1)  &  (0)& (6)   & (11)   & (5)   & \multicolumn{1}{|r}{(23)}  &  (49) & (836) \\
\multirow{2}{*}{SGR-MI}&  0.162 & 0.017 & 0.024     & 0.013 &  0.004 & \multicolumn{1}{|r}{0.057} & 0.219 & 0.114 & 0.022 & 0.002&  0.057& 0      & 0.018 & \multicolumn{1}{|r}{0.099} &  0.214&       \\
                       &   (88) &  (9) &  (13)     &  (7)  & (2)  & \multicolumn{1}{|r}{(31)}  &  (119)& (62)  &  (12) &  (1) & (31)  & (0)   & (10)   & \multicolumn{1}{|r}{(54)}  &  (116) & (543) \\

\addlinespace[5mm]   
\multicolumn{6}{l}{\colorbox{gray!30}{Multi-domain (AnnoMI)} }                & \multicolumn{1}{|r}{}      &       &       &       &      &       &       &       & \multicolumn{1}{|r}{}      &       &       \\
\multirow{2}{*}{MI-FS} &  0.026 & 0     & 0.002     & 0     & 0      & \multicolumn{1}{|r}{0.002} & 0.028 & 0.013 & 0.002 & 0    & 0.002 & 0.001 & 0     & \multicolumn{1}{|r}{0.004} & 0.017 &       \\
                       &   (30) &  (0)  &  (2)      &  (0)  & (0)    & \multicolumn{1}{|r}{(2)}   &  (32) & (15)  &  (2)  &  (0) & (2)   & (1)   & (0)   & \multicolumn{1}{|r}{(5)}   &  (20) & (1153)\\
\multirow{2}{*}{DIIR}  &  0.066 & 0.006 & 0.001     & 0.004 &  0.001 & \multicolumn{1}{|r}{0.012} & 0.078 & 0.047 & 0.002 & 0    &  0.009& 0.001 & 0.002 & \multicolumn{1}{|r}{0.015} &  0.062&       \\
                       &   (59) &  (5)  &  (1)      &  (4)  & (1)    & \multicolumn{1}{|r}{(11)}  &  (70) & (42)  &  (2)  &  (0) & (8)   & (1)   & (2)   & \multicolumn{1}{|r}{(13)}  &  (55) & (892) \\
\multirow{2}{*}{SGR-MI}&  0.151 & 0.023 & 0.026     & 0.033 &  0.018 & \multicolumn{1}{|r}{0.101} & 0.252 & 0.114 & 0.041 & 0    &  0.042& 0.002 & 0.009 & \multicolumn{1}{|r}{0.094} &  0.208&       \\
                       &   (82) &  (13) &  (14)     &  (18)  & (10)  & \multicolumn{1}{|r}{(55)}  &  (137)& (62)  &  (22) &  (0) & (23)  & (1)   & (5)   & \multicolumn{1}{|r}{(51)}  &  (113) & (543) \\

\addlinespace[5mm]   
\multicolumn{6}{l}{\colorbox{gray!30}{Combined (Diet-MI+AnnoMI)} }             & \multicolumn{1}{|r}{}      &       &       &       &      &       &       &       & \multicolumn{1}{|r}{}      &       &       \\
\multirow{2}{*}{MI-FS} &  0.040 & 0.002 & 0.002     & 0.001 & 0.000  & \multicolumn{1}{|r}{0.005} & 0.045 & 0.010 & 0.002 & 0    & 0.004 & 0.000 & 0     & \multicolumn{1}{|r}{0.006} & 0.016 &       \\
                       &   (88) &  (5)  &  (4)      &  (2)  & (1)    & \multicolumn{1}{|r}{(12)}   &  (100) & (21)  &  (5)  &  (0) & (8)   & (1)   & (0)   & \multicolumn{1}{|r}{(14)}   &  (35) & (2203)\\
\multirow{2}{*}{DIIR}  &  0.084 & 0.005 & 0.003     & 0.007 &  0.002 & \multicolumn{1}{|r}{0.017} & 0.101 & 0.039 & 0.002 & 0    &  0.008& 0.007 & 0.004 & \multicolumn{1}{|r}{0.021} &  0.060&       \\
                       &   (146) &  (9)  &  (5)      &  (12)  & (3)    & \multicolumn{1}{|r}{(29)}  &  (175) & (68)  &  (3)  &  (0) & (14)   & (12)   & (7)   & \multicolumn{1}{|r}{(36)}  &  (104) & (1728) \\
\multirow{2}{*}{SGR-MI}&  0.157 & 0.020 & 0.025     & 0.023 &  0.011 & \multicolumn{1}{|r}{0.079} & 0.236 & 0.114 & 0.031 & 0.001    &  0.050& 0.001 & 0.014 & \multicolumn{1}{|r}{0.097} &  0.211&       \\
                       &   (170) &  (22) &  (27)     &  (25)  & (12)  & \multicolumn{1}{|r}{(86)}  &  (256)& (124)  &  (34) &  (1) & (54)  & (1)   & (15)   & \multicolumn{1}{|r}{(105)}  &  (229) & (1086) \\                       

\bottomrule
\end{tabular}

}

\caption{Proportion of occurrences of the question categories defined in Table \ref{tab:app_q_category_def} to the number of full sentences in the system utterances. Numbers in parentheses represent counts. ev-* is the sum of ev-d, ev-a, ev-r, ev-n. ev* is the sum of ev and ev-*. pl-* is the sum of pl-a, pl-c, pl-d, pl-t, pl-p. pl* is the sum of pl and pl-*.}
\label{tab:app_q_category_full}
\end{table*}

\begin{table*}[]
\centering
\resizebox{1\linewidth}{!}{%
\begin{tabular}{p{10mm}p{180mm}}
\toprule
ID	&	Utterance	\\ 
\midrule
1-Cl	&	Hello. I'm looking for some advice to help improve my diet. Since having my son I tend to snack a lot and these are generally sugary snacks. I eat  meals but need advice on how to make these meals more filling so I don't need to snack. \\
2-Co    & It's great that you're looking to make changes to your diet. Perhaps incorporating more protein and fiber-rich foods into your meals could help you feel fuller for longer and reduce the need for sugary snacks. \textcolor{blue}{(The system tends to jump to offering solutions instead of delving deeper into the context of the user's issue.)}\\
3-Cl    & Can we go through it meal by meal? When I wake up I always have a coffee and a bowl of Fruit n Fibre cereal but by the time I've done the 2 mile walk back and forth from school, I'm hungry again \\
4-Co   & That sounds like a busy start to your day. Perhaps adding a source of protein to your breakfast, like a boiled egg or some Greek yogurt, might help keep you full until lunch. \\
5-Cl   & I could try a boiled egg. How about lunch? I generally have a plain bagel because I'm short on time but this isn't filling or healthy. What suggestions can you make? \\
6-Co   & You might try adding some lean protein like turkey or chicken breast to your bagel, along with some veggies or avocado to boost the nutritional value and make it more satisfying. Alternatively, you could prepare a quick and easy salad with a protein source the night before to save time. \\
7-Cl   & Meal prepping for lunch sounds like a great idea. Can you suggest salad ideas with protein? \textcolor{blue}{(User increasingly seeks advice from the system because it offers informative suggestions.)} \\
8-Co   & Certainly! You could try a spinach and grilled chicken salad with cherry tomatoes, cucumbers, and a sprinkle of feta cheese, or a quinoa salad with chickpeas, roasted vegetables, and a lemon-tahini dressing for a vegetarian option. \\
9-Cl   & How about dinner? I generally have a decent meal and don't tend to snack after it but I feel like these meals could be healthier as sometimes I opt for easy meals like pizza or a microwave meal to save time \\
10-Co  & For dinner, you might consider incorporating more whole foods by preparing simple yet nutritious meals like a stir-fry with lots of vegetables and a protein like tofu or chicken. You can also make extra servings and use leftovers for a quick lunch the next day, combining convenience with healthy eating.\\ 
11-Cl  & As I am single I do get bored with eating the same things day to day because every meal kit etc tends to be for 2-4 people so then have to have the same things again \\
12-Co  & You might try experimenting with different herbs and spices to add variety to your meals, or look into single-serving recipes online that cater to a variety of tastes. Rotating through a set of diverse protein and vegetable options can also keep things interesting while maintaining a balanced diet. \\
13-Cl  & If I do want a snack, what healthy options can you recommend instead of crips, chocolate biscuits etc \\
14-Co  & You could try snacking on raw vegetables with hummus, a handful of nuts for a protein boost, or some fresh fruit which can satisfy your sweet cravings while providing vital nutrients. Greek yogurt with berries or a small piece of dark chocolate might also be good alternatives for a healthy snack. \\
15-Cl        & I do like snacking on dates but again tend to opt for convenient options rather than spending time peeling a carrot \\
16-Co        & That's understandable. Dates are a great choice as they're naturally sweet and nutritious, so keeping them on hand for convenience can be helpful. Additionally, you could consider pre-cutting vegetables like carrots or celery sticks to have ready for a quick and easy snack. \textcolor{blue}{(The system proactively provides information without being explicitly asked for advice.)}\\
17-Cl        & I don't like celery, do you have other suggestions \\
18-Co        & Of course! Aside from carrots, you might try bell peppers, cherry tomatoes, or cucumber slices as easy snack options that can also be pre-prepared for convenience. \\
19-Cl        & Can you give me a menu idea for 1 day that would help fill me up? \\
20-Co        & Sure! For breakfast, you could have oatmeal topped with mixed berries and a sprinkle of nuts. For lunch, enjoy a whole grain wrap with turkey, spinach, avocado, and a side of baby carrots. For dinner, try a grilled salmon filet with quinoa and roasted broccoli, and if you need a snack, you can have Greek yogurt with a drizzle of honey. \\
21-Cl        & I tend to find if I do more in a day like 5km on the exercise bike I tend to get very hungry \\
     &	... (Subsequent utterances were omitted.)	\\
\bottomrule
\end{tabular}
}

\caption{Example of dialogue with MI Few-shot (MI-FS) dialogue system. `Cl' denotes Client (user) and `Co' denotes Counselor (system). \textcolor{blue}{Blue letters in parentheses indicate comments on the corresponding utterance.}}
\label{tab:sample_MI-FS}

\end{table*}
\begin{table*}[]
\centering
\resizebox{1\linewidth}{!}{%

\begin{tabular}{p{10mm}p{180mm}}
\toprule

ID	&	Utterance	\\
\midrule

1-Cl	&	Hi, I am struggling with my diet \\
2-Co & It sounds like you're feeling torn about how to manage your diet right now.\\
3-Cl & I really am, I am trying to restrict my calories but cannot seem to stop snacking \\
4-Co &How about trying to keep a small journal to track when and what you snack on, so we can identify any patterns and make small adjustments? \textcolor{blue}{(The system proposes ideas without an explicit request from the user.)}\\
5-Cl & I already track my food but i haven't thought about keeping time diaries, how will this help? \\
6-Co & A time diary can help pinpoint specific times or situations that trigger your snacking habits. This information can be useful in understanding your behavior patterns more clearly. \\
7-Cl & If i say identfy when it is happening, hgow can i stop it? \\
8-Co & Try having a glass of water or taking a short walk when you feel the urge to snack, as this can help break the habit and give you a moment to reassess your hunger. \textcolor{blue}{(In response to user requests, the system often offers information and advice.)}\\
9-Cl & Can you suggest a 2000 calorie diet that is balanced so that i stop snacking? \\
10-Co & It sounds like you're open to exploring different strategies to manage your snacking, including tracking your eating patterns and considering a balanced diet. Acknowledging these steps is a great start, and working with a nutritionist can help tailor a plan that fits your needs. \\
11-Cl & The problem is, i am on a budget right now \\
12-Co & It sounds like you're feeling a bit overwhelmed by trying to balance your dietary goals with financial constraints. It's completely understandable to feel this way, and I'm here to support you as we navigate these challenges together. \\
13-Cl & can you suggest ways i could eat well on a budget? \\
14-Co & It's great to see your motivation to eat well on a budget, and focusing on affordable staples and meal planning is a solid approach. Remember, small, consistent changes can make a big difference over time. \\
15-Cl & what is a good amount of protein for a 185lb male in his 30s? \\
16-Co & It's wonderful to see your proactive approach in considering how to maintain a healthy diet within your budget, and your openness to exploring these changes is truly commendable. \\
17-Cl & can you suggest ways i can be pro active? \\
18-Co & What are some specific areas in your life where you feel you could take more initiative or make positive changes? \\
19-Cl & with fitness \\
20-Co & What aspects of your current fitness routine do you feel most excited about improving or changing? \\
21-Cl & my cardio routine \\
22-Co & It sounds like you're eager to enhance your cardio routine and are open to exploring different options to make it more effective. By focusing on this area, you're taking a proactive step towards improving your fitness, which is a great mindset to have. \\
23-Cl & What are some examples of cardio i can do? \\
24-Co & What types of activities have you enjoyed in the past that might inspire your cardio routine now? How do you envision these activities fitting into your current lifestyle? \\
25-Cl & I have enjoyed hiking in the past \\
26-Co & It sounds like you're considering incorporating hiking into your cardio routine, drawing from an activity you've enjoyed in the past. This shows you're thinking about how to blend fitness with your interests. \textcolor{blue}{(Although the inclusion of meaning-adding responses resulted in a high rate of complex reflections in DIIR, these reflections fell short of effectively eliciting the user's thoughts (similar to ID 22-Co).)}\\
27-Cl & I am indeed, are there any things that are complimentary to hiking that would help my fitness? \\
28-Co & It's great to see you exploring complementary activities to hiking, as this can enhance your overall fitness and enjoyment. By considering options like strength training or flexibility exercises, you're building on your past experiences and interests to create a well-rounded routine. \\
	&	... (Subsequent utterances were omitted.)	\\

\bottomrule
\end{tabular}
}

\caption{Example of dialogue with the DIIR dialogue system. `Cl' denotes Client (user) and `Co' denotes Counselor (system). \textcolor{blue}{Blue letters in parentheses indicate comments on the corresponding utterance.}}
\label{tab:sample_MI-DIIR}

\end{table*}
\begin{table*}[]
\centering
\resizebox{1\linewidth}{!}{%

\begin{tabular}{p{10mm}p{180mm}}
\toprule

ID	&	Utterance	\\
\midrule
1-Cl	&	Hi there, I'll like to discuss a personal issue I'll like to address. can you help me with that? \\
2-Co & Of course, I'm here to help. Could you share some details about the issue you're facing? \\
3-Cl & I'm trying to improve my speech (especially conversational) and I find it hard to practice doing this regularly \\
4-Co & Could you tell me more about what makes it difficult for you to practice your speech regularly? \textcolor{blue}{(Ask about the importance of the issue.)}\\
5-Cl & I think it's down to my schedule and environment. I'm self-employed and my work takes most of my day. Second, I live with another person who might be disturbed by my practice whenever they're around. Although, I believe that I can always make up time, but I just can't with these two factors also being the major friction \\
6-Co & What are some specific strategies you believe could help you manage your schedule or create a more suitable environment for practicing your speech? \\
    & ... (Intermediate utterances omitted) ...\\
15-Cl & Because of my history of inconsistency. I'm hoping you'll give me some strategies to imbibe this habit - consistently. \\
16-Co & What strategies do you think might help you become more consistent with your practice schedule, despite the challenges with your schedule and environment? \textcolor{blue}{(Prioritize questions that elicit the user's own ideas, even when the client explicitly requests advice.)}\\
17-Cl & Like I mentioned earlier, dedicating blocks of time to practice and research social interaction. For example, regular 5 to 10 minutes during the day. \\
18-Co & You're thinking of integrating regular 5 to 10 minutes of practice during the day as a strategy to build consistency with your practice schedule. \\
19-Cl & Absolutely. What do you think and how can I make this realistic? \\
20-Co & It sounds like you are considering incorporating regular 5 to 10 minute practice sessions into your day to help build consistency in your practice schedule. \textcolor{blue}{(Although the user requests advice, the proposed method demonstrates a tendency to provide reflections on the user's spoken content instead of offering advice.)}\\
21-Cl & Yes. That's right. What do you think about this approach? \\
22-Co & How confident are you in your ability to consistently implement these regular 5 to 10 minute practice sessions during your day? \\
23-Cl & Unfortunately, not very confident. \\
24-Co & It sounds like you're not very confident in maintaining the consistency needed for the 5 to 10 minute practice sessions during your day. \\
25-Cl & No I'm not. How can you help me with this? \\
26-Co & What might increase your confidence in being able to consistently implement these 5 to 10 minute practice sessions? \\
27-Cl & I'm thinking some form of motivational triggers or strategies \\
28-Co& What types of motivational triggers or strategies do you think would make you feel more confident in maintaining the consistency for your practice sessions? \\
29-Cl & Maybe rewards for practising or potential consequences of not practicing. What ideas can you give me? \textcolor{blue}{(By asking questions to prompt the user's thoughts, the system succeeds in extracting ideas from the user.)} \\
30-Co & It sounds like you're considering implementing motivational triggers or strategies, such as rewards or recognizing consequences, to help with maintaining consistency in your practice sessions. \\
31-Cl & Absolutely. Can you give some ideas of what that my entail (practically) for my situation - so I can consistently implement my strategies to improve my speech? \\
32-Co & It sounds like you're considering practical motivational triggers, like rewards for practicing or recognizing the potential consequences of not practicing, as strategies to increase your consistency. \\
33-Cl & Yes I am. Can you give some ideas on how to do this? \\
34-Co & One practical idea for implementing motivational strategies could be setting small rewards for each completed practice session, like treating yourself to a favorite activity, or acknowledging the positive outcomes of staying consistent to reinforce progress. \\
35-Cl & Can you give me some practical examples of that? \\
36-Co & What specific rewards or consequences do you think would be most motivating and practical for you to implement in your daily routine? \\
37-Cl & Perhaps vividly envisioning how improving my speech can transform my life - socially, financially etc., and also how to improving it might stop me from making these happen. I might also look at things I enjoy doing during the day, but won't do them until I complete the block of practice preceding them. What do you think? \textcolor{blue}{(Prompting the user to articulate specific ideas about rewards led to the extraction of a concrete plan.)}\\
38-Co & It sounds like you're really considering some practical strategies to motivate yourself, such as envisioning the benefits of improving your speech and using activities you enjoy as rewards for practicing consistently. These seem like creative ways to help maintain your routine. \\
	&	... (Subsequent utterances were omitted.) \\
\bottomrule 
\end{tabular}

}

\caption{Example of dialogue with the proposed system (SGR-MI). `Cl' denotes Client (user) and `Co' denotes Counselor (system). \textcolor{blue}{Blue letters in parentheses indicate comments on the corresponding utterance.}}
\label{tab:sample_ours}

\end{table*}

\end{document}